\documentclass[12pt,preprint]{aastex}
\usepackage{amsmath}

\let\url\relax
\usepackage[hypertex]{hyperref}
\makeatletter
\renewcommand{\fps@figure}{tp}
\makeatother
\setkeys{Gin}{width=\linewidth,height=0.9\textheight,keepaspectratio=true}

\newcommand{\TA}{\tablenotemark{a}}
\newcommand{\TB}{\tablenotemark{b}}
\newcommand{\TC}{\tablenotemark{c}}
\newcommand{\TD}{\tablenotemark{d}}
\newcommand{\TE}{\tablenotemark{e}}
\newcommand{\TF}{\tablenotemark{f}}
\newcommand{\TG}{\tablenotemark{g}}
\newcommand\nd{\nodata}

\newcommand{\elecd}{$n_{\rm e}$}
\newcommand{\elect}{$T_{\rm e}$}

\newcommand{\oiii}{\ion{O}{3}}
\newcommand{\nitroi}{\ion{N}{1}}
\newcommand{\nii}{\ion{N}{2}}
\newcommand{\niii}{\ion{N}{3}}

\newcommand{\silii}{\ion{Si}{2}}
\newcommand{\oi}{\ion{O}{1}}
\newcommand{\oii}{\ion{O}{2}}

\newcommand{\cii}{\ion{C}{2}}

\newcommand{\sii}{\ion{S}{2}}

\newcommand{\ciii}{\ion{C}{3}}
\newcommand{\cai}{\ion{Ca}{1}}

\newcommand{\cliii}{\ion{Cl}{3}}

\newcommand{\feii}{\ion{Fe}{2}}
\newcommand{\feiii}{\ion{Fe}{3}}
\newcommand{\feiv}{\ion{Fe}{4}}

\newcommand{\ariv}{\ion{Ar}{4}}
\newcommand{\hi}{\ion{H}{1}}
\newcommand{\hii}{\ion{H}{2}}

\newcommand{\hei}{\ion{He}{1}}
\newcommand{\heii}{\ion{He}{2}}

\newcommand{\tmthree}{10$^{-3}$}

\newcommand{\ts}{\emph{$t^{\rm 2}$}}

\shorttitle{\cii and \oii Lines in EHRs}
\shortauthors{Esteban et al.}

\received{}
\begin{document}

\title{Keck HIRES Spectroscopy of Extragalactic \hii\ Regions: C and O Abundances from Recombination Lines\footnotemark{}}

\author{C\'esar Esteban}
\affil{Instituto de Astrof{\'\i}sica de Canarias, E-38200 La Laguna, Tenerife, Spain; and Departamento de Astrof{\'\i}sica, 
Universidad de La Laguna, E-38071 La Laguna, Tenerife, Spain}
\email{cel@iac.es}

\author{Fabio Bresolin}
\affil{Institute for Astronomy, 2680 Woodlawn Drive, Honolulu, HI 96822, USA}
\email{bresolin@ifa.hawaii.edu}

\author{Manuel Peimbert}
\affil{Instituto de Astronom\'\i a, UNAM, Apdo. Postal 70-264, M\'exico 04510 D.F., Mexico}
\email{peimbert@astroscu.unam.mx}

\author{Jorge Garc\'{\i}a-Rojas}
\affil{Instituto de Astronom\'\i a, UNAM, Apdo. Postal 70-264, M\'exico 04510 D.F., Mexico}
\email{jgarcia@astroscu.unam.mx}

\author{Antonio Peimbert}
\affil{Instituto de Astronom\'\i a, UNAM, Apdo. Postal 70-264, M\'exico 04510 D.F., Mexico}
\email{peimbert@astroscu.unam.mx}

\author{Adal Mesa-Delgado}
\affil{Instituto de Astrof{\'\i}sica de Canarias, E-38200 La Laguna, Tenerife, Spain}
\email{amd@iac.es}

\begin{abstract}
We present very deep spectrophotometry of 14 bright extragalactic \hii\ regions belonging to spiral, irregular, and blue 
compact galaxies. 
The data for 13 objects were taken with the HIRES echelle spectrograph on the Keck I telescope. We have measured \cii\ recombination lines in 10 of the objects and \oii\ recombination lines in 8 of them. We have determined electron temperatures from line ratios 
of several ions, specially of low ionization potential ones. We have found a rather tight linear empirical relation between {\elect}([{\nii}]) and {\elect}([{\oiii}]). We have 
found that \oii\ lines give always larger abundances than [{\oiii}] lines. Moreover, the difference of both O$^{++}$ abundance determinations  --the so-called abundance discrepancy factor-- is very similar in all the objects, with a mean value of 0.26 $\pm$ 0.09 dex, independently of the properties of the \hii\ region and of the parent galaxy. Using the observed recombination lines, we have determined the O, C, and C/O 
radial abundance gradients for 3 spiral galaxies: M~33, M~101, and NGC~2403, finding that C abundance gradients are always steeper than those of O, producing negative C/O gradients accross the galactic disks. This result is similar to that found in the Milky Way 
and has important implications for chemical evolution models and the nucleosynthesis of C.
\end{abstract}

\keywords{galaxies: abundances --galaxies: individual (M~31, M~33, M~83, M~101, NGC~1741, NGC~2366, NGC~2403, NGC~4395, NGC~4861) -- galaxies: ISM -- galaxies: spiral -- {\hii} regions}

\section{Introduction}

\footnotetext{Most of the data presented herein were obtained at the W.M. Keck Observatory, which is operated as a scientific partnership among the California Institute of Technology, the University of California and the National Aeronautics and Space Administration. The Observatory was made possible by the generous financial support of the W.M. Keck Foundation. Part of the observations were made with the 4.2 m William Herschel Telescope (WHT), operated on the island of La Palma by the Isaac Newton Group in the Spanish Observatorio del Roque de los Muchachos of the Instituto de Astrof\'\i sica 
de Canarias.}

\label{intro}

The spectral analysis of {\hii} regions allows the determination of  the chemical composition of the ionized gas-phase of the 
interstellar medium from the solar neighbourhood to the high-redshift Universe. Therefore, it stands as an essential tool 
for our knowledge of the chemical evolution of the Universe. With the advent of 8-10 m class ground-based telescopes, 
we can now obtain extremely deep spectra of extragalactic {\hii} regions (hereafter EHRs).  
These new studies have permitted, for example, to obtain direct determinations of the electron temperature, {\elect}, 
in high-metallicity EHRs \citep[see][and references therein]{bresolin08}, where the auroral lines 
become very faint, 
or to measure recombination lines (hereafter RLs) useful for abundance determinations of heavy-element ions \citep{peimbert03,lopez-sanchezetal07,bresolin07}. 

The detection of {\cii} and {\oii} lines produced by pure recombination in EHRs was firstly reported by 
\citet{estebanetal02} from deep spectra taken with the 4.2 m William Herschel Telescope. In principle, these lines 
have the advantage that their intensity is much less dependent on the value of {\elect} than 
the collisionally excited lines (hereafter CELs), which are the lines commonly used for abundance determinations 
in nebulae. 
The brightest {\cii} RL is {\cii} $\lambda$4267, with typical fluxes of the order of {\tmthree} $\times$ $I$(H$\beta$). 
This line permits to derive the C$^{++}$ abundance, which is the dominant ionization stage of C for the typical conditions 
of EHRs. There are only a few C abundance determinations available for EHRs, most of 
them derived from UV CELs that can only be observed from space \citep{garnettetal95,garnettetal99,kobulnickyskillman98}, and  more recently from RLs \citep{estebanetal02,peimbert03,tsamisetal03,peimbertetal05,lopez-sanchezetal07,bresolin07}. The C abundance determinations 
based on UV CELs are severely affected by uncertainties in the reddening correction. To further complicate the situation, the STIS spectrograph aboard the {\it Hubble Space Telescope}, the only instrument capable to detect the UV CELs of C in bright EHRs, stopped science operations 
in 2004, so that nowadays the observation of the optical CII RLs provides the only possibility for 
determining C abundances in EHRs. The study of the behavior of 
C/H and C/O ratios and their galactocentric gradients in galaxies of different morphological types and  
metallicities is of paramount importance and can provide observational constraints for a better knowledge of 
the nucleosynthetic origin of carbon --the most important biogenic element-- as well as the star formation/enrichment timescales in galaxies \citep[e.g.][]{carigietal05}.\\ 

O$^{++}$ is the only ion that simultaneously shows bright CELs and detectable --relatively bright-- RLs in the optical range. In {\hii} regions, 
the O$^{++}$ abundances derived from RLs are always between 0.10 and 0.35 dex higher than those derived from CELs 
\citep[see compilation by][]{garcia-rojasesteban07}. This observational fact is 
currently known as the ``abundance discrepancy" (hereafter AD) problem. We define the abundance discrepancy factor (hereafter 
ADF) as the logarithmic difference between the abundance derived from RLs and CELs: 
\begin{eqnarray}
 {\rm ADF}({\rm X}^{i+}) = {\rm log}({\rm X}^{i+}/{\rm H}^+)_{\rm RLs} - {\rm log}({\rm X}^{i+}/{\rm H}^+)_{\rm CELs} ,
\end{eqnarray}
where X$^{i+}$ corresponds to the ionization state $i$ of element X. Although the ADF found in {\hii} regions is remarkably constant, 
this is not the case for planetary nebulae, PNe, where the values of the ADF can be quite different from object 
to object, with the ADFs ranging from near zero up to ten in some extreme cases \citep[e.g.][]{liu06}. This different behaviour  
led \citet{garcia-rojasesteban07} to the conclusion that the mechanism that produces the AD --or the bulk of it-- in the extreme PNe should be different to 
that producing the AD in {\hii} regions. 
The fact that the determination of nebular abundances is still uncertain, by at least  a factor of two,  has an important impact on many 
astrophysical aspects, such as chemical evolution and nucleosynthesis models and predictions, as well as the calibration 
of the so-called ``strong line methods" used to estimate abundances in local and high-redshift star-forming galaxies. This last point was recently highlighted by \citet{peimbertetal07}. 
  
The origin of the AD problem is still object of debate and a challenge for our understanding of ionized nebulae. The results for a sample of Galactic {\hii} regions seem 
to be consistent with the predictions of the temperature fluctuations paradigm \citep{garcia-rojasesteban07}. In fact, 
in the presence of temperature fluctuations (parameterized by the mean square of the spatial variations of temperature, 
the \ts\ parameter) the AD can be naturally explained because of the different temperature dependence of the intensity of RLs and CELs. However, the presence of temperature fluctuations in ionized nebulae still lacks a direct demonstration. An alternative explanation for the origin of the AD has been proposed by \citet{tsamispequignot05} and \citet{stasinskaetal07}, which is based on the presence of cold high-metallicity clumps of supernova ejecta still not 
mixed with the ambient gas of the {\hii} regions. This cold gas would produce most of the emission of the RLs whereas 
the ambient gas of normal abundances would emit most of the intensity of CELs. However, \citet{lopez-sanchezetal07},
who detect {\cii} and {\oii} lines in the dwarf galaxy NGC~5253, question this hypothesis in the light of the results 
available for EHRs.  

Our group is interested in exploring the physical process that control the AD as 
well as in determining C abundances and radial gradients. 
In this paper, we make use of deep high-resolution spectrophotometry of a sample of bright EHRs in spiral and irregular galaxies as well as some giant \hii\ regions in dwarf galaxies --\hii\ galaxies-- in order to explore a wide range of 
metallicities, \hii\ regions of different sizes and structure, and galaxy morphological types. 
We think that this kind of study  is necessary, mainly because the only intermediate-resolution spectral 
study devoted for the detection of RLs in EHRs of spiral galaxies was that by \citet{estebanetal02}, which only included 
--apart from NGC~2363 in the irregular galaxy NGC~2366-- three objects in two spirals (one in M~33 and two in M~101). 

A highlight of this work is the use of high-spectral resolution spectrophotometry. These data are needed 
to increase the contrast between the faint RLs and the continuum, to deblend the \oii\ lines of multiplet 1, and to separate 
them properly from possible Wolf-Rayet emission features as well as absorption features due to underlying stellar populations.  

In \S\S~\ref{obsred} and~\ref{lineint} of this paper we describe the observations, the data reduction procedure, 
the measurement of the emission lines, and the derivation of the reddening coefficient. In \S~\ref{phiscondabund} we 
derive the physical conditions of the nebulae and explore the consistency of different temperature scales, 
as well as calculate the ionic abundances from both kinds of lines: CELs and RLs. In \S~\ref{discussion} we discuss the ADF in the 
sample objects and the radial C, O, and C/O abundance gradients in the spiral galaxies M~33, M~101, and NGC~2403. Finally, in \S~\ref{conclu} we summarize our main conclusions. 

\section{Observations and Data Reduction}
\label{obsred}
The observations of the sample objects, except NGC~5447 --a bright \hii\ region in the galaxy M~101-- were made on 2006 April 20 and 21 and November 14 with the High Resolution Echelle Spectrometer \citep[HIRES,][]{vogtetal94} 
at the Keck I telescope on Mauna Kea Observatory. The spectra cover the 3550--7440 \AA\ range with a somewhat discontinuous wavelength coverage due to gaps between the detectors and to the fact that the redmost spectral orders do not fit completely within the CCD. The decker D3 was used, covering an area of 7\farcs0 $\times$ 1\farcs7. This configuration provides a spectral resolution of R=23,000. 
We observed a single slit position for each object. The center and position angle of the slits were chosen to cover the maximum 
area of the brightest part of the objects and the final usable one-dimensional spectra were extracted from an area of 5\farcs76 $\times$ 1\farcs7
for all the objects. The journal of the observations is shown in Table~\ref{observations}. The table includes the coordinates of the centers of 
the slits. In the case of the \hii\ regions belonging to spiral galaxies with two or more objects 
observed --M33, NGC~2403, and M101, all of morphological type Sc-- we also include the galactocentric distance (in kpc) and $R$/$R_{0}$ ratio of each \hii\ region. The adopted distances to the 
galaxies have been taken from \citet{freedmanetal01}, and are 0.84, 3.22, and 7.5 Mpc for M33, NGC~2403, 
and M101, respectively. For M33, we have considered the photometric radius, $R_{0}$, of 28\arcmin\ estimated by \citet{devaucouleursetal76}, an inclination 
angle of 56$^{\circ}$, and a position angle of 23$^{\circ}$ \citep{zaritskyetal89}. In the case of NGC~2403, we have adopted $R_0$ = 8\farcm5 \citep{zaritskyetal94}, $i$ = 60$^{\circ}$, and PA = 126$^{\circ}$ \citep{garnettetal97}. Finally, for M101 we have assumed $R_0$ = 14\farcm4 
\citep{devaucouleursetal91}, $i$ = 18$^{\circ}$, and PA = 37$^{\circ}$ \citep{kamphuis93}.

Intermediate-resolution spectroscopy of NGC~5447 was obtained on 2008 May 11 with the ISIS spectrograph  
at the 4.2m William Herschel Telescope (WHT) of the Observatorio del Roque de los Muchachos 
(La Palma, Spain). Two different CCDs were used at the blue and red arms of the spectrograph: 
an EEV CCD with a configuration 4100 $\times$ 2048 pixels with a pixel size of 13.5 $\mu$m in the blue arm and a 
REDPLUS CCD with 4096 $\times$ 2048 pixels with a pixel size of 15 $\mu$m in the red arm. The dichroic prism used to separate 
the blue and red beams was set at 5300 \AA. The slit was 3\farcm7 long and 1\farcs03 wide. Two gratings 
were used, the R1200B in the blue arm and the R316R in the red arm. These gratings give reciprocal 
dispersions of 17 and 62 \AA\ mm$^{-1}$, and effective spectral resolutions of 0.86 and 3.56 \AA\ 
for the blue and red arms, respectively. The blue spectra cover from $\lambda\lambda$4225 to 5075 \AA\ and the red 
ones from $\lambda\lambda$5430 to 8195 \AA. The one-dimensional spectra were extracted from an area of 3\farcs2 $\times$ 1\farcs03, 
corresponding to the brighest zone of NGC~5447. 

Standard data reduction procedures, including bias correction, flat-fielding, order extraction, wavelength calibrations, and flux calibration, were carried out using 
routines in the ECHELLE and ONEDSPEC packages of IRAF\footnote{IRAF is distributed by National Optical Astronomical Observatories, operated by the Associated Universities for Research in Astronomy, Inc., under contract to the National Science Foundation}. 
The correction for atmospheric extinction was performed using the 
average curve for continuous atmospheric extinction at Mauna Kea and Roque de los Muchachos. The absolute flux calibration 
was achieved by observations of the standard stars Feige~34, Feige~66, Feige~110, H600, BD+25~4655, and BD+28~4211 in the case of the 
observations taken with the Keck I telescope and Feige~34, BD+25~3941, and BD+33~2624 in the case of the observations taken with the WHT. 

\begin{deluxetable}{lccccccccc}
\tabletypesize{\scriptsize}
\rotate
\tablecaption{Journal of observations
\label{observations}}
\tablewidth{0pt}
\tablehead{
& & & & & & & & & 
\colhead{Exposure} \\
\colhead{Host}  & 
 & 
\colhead{R.A.{\TA}}  & 
\colhead{Decl.{\TA}} &
\colhead{$R_{\rm G}$} &
&
\colhead{Date of} &
&
\colhead{P.A.}  & 
\colhead{Time} \\ 
\colhead{Galaxy} &
\colhead{Object}  &   
\colhead{(J2000.0)} & 
\colhead{(J2000.0)} & 
\colhead{(kpc)} &
\colhead{$R$/$R_0$} &
\colhead{Observation} &
\colhead{Telescope} &
\colhead{(deg)} &
\colhead{(s)}}  
\startdata 
M31 & K932{\TB} & 00 46 34.0 & +42 11 51 & \nd & \nd & 14/11/2006 & Keck & 330 & 3$\times$1800 \\
M33 & NGC~595 & 01 33 34.2 & +30 41 38 & 2.87 & 0.42 & 14/11/2006 & Keck & 90 & 3$\times$1800 \\
M33 & NGC~604 & 01 34 33.3 & +30 46 47 & 4.11 & 0.60 & 14/11/2006 & Keck & 90 & 3$\times$1200 \\
NGC~1741 & Zone C{\TC} & 05 01 37.6 & $-$04 15 30 & \nd & \nd & 14/11/2006 & Keck & 50 & 2$\times$1200 \\
NGC~2366 & NGC~2363 & 07 28 42.7 & +69 11 26 & \nd & \nd & 20/04/2006 & Keck & 220 & 3$\times$1800 \\
NGC~2403 & VS~24{\TD} & 07 36 45.7 & +65 36 55 & 1.00 & 0.126 & 14/11/2006 & Keck & 295 & 3$\times$1500 \\
NGC~2403 & VS~38{\TD} & 07 36 51.8 & +65 36 42 & 1.04 & 0.131 & 14/11/2006 & Keck & 275 & 3$\times$1650 \\
NGC~2403 & VS~44{\TD} & 07 37 06.8 & +65 36 41 & 2.77 & 0.346 & 14/11/2006 & Keck & 330 & 3$\times$1500 \\
NGC~4395 & Region \#70{\TE} & 12 25 57.7 & +33 31 32 & \nd & \nd & 21/04/2006 & Keck & 330 & 3$\times$1800 \\
NGC~4861 & Brightest HII region{\TF} & 12 59 00.1 & +34 50 37 & \nd & \nd & 20/04/2006 & Keck & 220 & 3$\times$1800 \\
M83 & Nucleus & 13 36 59.8 & $-$29 52 07 & \nd & \nd & 21/04/2006 & Keck & 90 & 3$\times$1800 \\
M101 & H1013{\TG} & 14 03 32.2 & +54 21 10 & 5.50 & 0.19 & 20/04/2006 & Keck & 270 & 4$\times$1800 \\
M101 & NGC~5461 & 14 03 41.0 & +54 19 01 & 9.84 &  0.34 & 20/04/2006 & Keck & 225 & 3$\times$1800 \\
M101 & NGC~5447 & 14 02 28.7 & +54 16 25 & 16.21 & 0.56 & 11/05/2008 & WHT & 134 & 7$\times$1800 \\
\enddata
\tablenotetext{a}{Coordinates of the slit center. Units of right ascension are hours, minutes, and seconds, and units of declination are degrees, arcminutes, and arcseconds. }
\tablenotetext{b}{Walterbos \& Braun (1992) and Galarza et al. (1999)}
\tablenotetext{c}{Hickson (1982)}
\tablenotetext{d}{V\'eron \& Sauvayre (1965)}
\tablenotetext{e}{Cedr\'es \& Cepa (2002)}
\tablenotetext{f}{Gil de Paz et al. (2003)}
\tablenotetext{g}{Hodge et al. (1990)}
\end{deluxetable}

\section{Line intensities and Reddening Correction}  
\label{lineint}

Line intensities were measured integrating all the flux in the line between two given limits and 
over a local continuum estimated by eye. In the cases of line blending, a multiple Gaussian profile fit procedure was 
applied to obtain the line flux of each individual line. The measurements were performed with the SPLOT routine of 
the IRAF package.

\begin{figure*}
  \includegraphics[angle=0,scale=1]{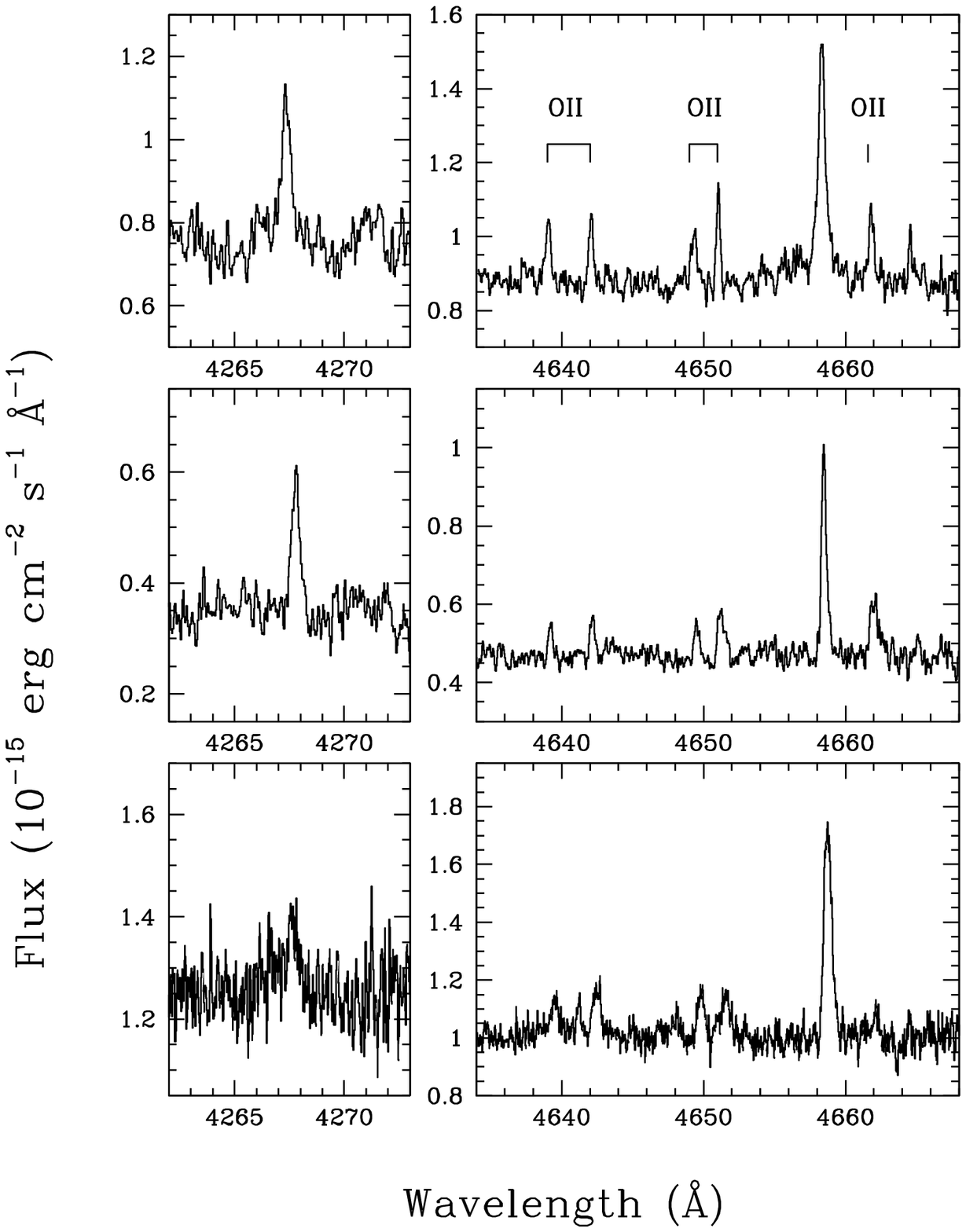} 
  \caption{Sections of the spectra of NGC~604 (upper panels), K932 (middle panels), and NGC~2363 (lower panels) showing 
  the \cii\ $\lambda$4267 (left) recombination line and the lines of multiplet 1 of \oii\ $\sim$$\lambda$$\lambda$4650 (right).}
  \label{RLs}
\end{figure*}

Table~\ref{lineid1} and Table~\ref{lineid2} show the emission-line intensities measured for 10 EHRs where the RLs of 
{\oii} and/or {\cii} have been detected. 
Table~\ref{lineid3} includes the emission-line intensities for 4 objects where those RLs were not detected. 
Due to their relevance to this paper, upper limits of the intensities of {\cii} $\lambda$4267 and/or \oii\ $\lambda$4649 
--one of the brightest lines of multiple 1 of {\oii}-- at 1 $\sigma$ level have been estimated and 
included in the tables for the objects where those RLs were not detected. 
For M~83, we cannot estimate appropriate upper limits of the intensities of {\cii} and {\oii} lines due to the contamination of spectral features  
produced by the underlying stellar emission to the continuum at low intensities. 
The first column of tables~\ref{lineid1} to \ref{lineid3} includes the adopted laboratory wavelength, $\lambda_0$, of the line. The second and 
third columns include the ion and identification of the line. The fourth column lists the reddening curve used \citep{seaton79}. 
The following columns include --one pair of columns for each object-- the observed wavelength in the heliocentric rest frame, 
$\lambda$, and the reddening-corrected flux relative to H$\beta$, $I$($\lambda$), of the lines. 
The identification and adopted laboratory wavelength of the lines, as well as the error analysis applied, were obtained following \citet{garcia-rojasetal04}, adding quadratically the error due to flux calibration that has been estimated to 
be 1 and 3 per cent for the November and April 2006 Keck I observations, respectively, and 4 per cent for the WHT observations. 
This flux calibration error corresponds to the standard deviation obtained from the calibration curves of the standard stars observed in each run. 
Colons indicate line flux errors of the order or greater than 40\%. 

A number of lines between 38 (M83-Nucleus) and 135 (NGC~2363) have been measured in the spectra of the 14 objects 
included in this study. A substantial fraction of them are permitted lines of heavy-element ions: \cii, \nitroi, \oi, \oii, and \silii; of these, only 
\cii\ $\lambda$4267 and the lines of multiplet 1 of \oii\ at $\sim$$\lambda\lambda$4650 are pure RLs and can be used 
for abundance determinations. The rest of the permitted lines are mainly produced by fluorescence mechanisms 
\citep[see][]{estebanetal04}. Only a small fraction of the lines are dubious identifications or could not be identified. 
Figure~\ref{RLs} contains sections of the spectra of three of the sample objects --NGC~604, K932 and NGC~2363-- showing  
the \cii\ $\lambda$4267 line and the lines of multiplet 1 of \oii. 

All the sample objects have optical spectra published in the literature. In the case of the \hii\ region K932 in M~31, only the 
intensities of a few bright emission lines have been reported \citep{galarzaetal99}, no electron temperatures nor direct determinations of the chemical abundances were available for this object. This is the first time that \cii\ and 
\oii\ lines are detected in this object and in an \hii\ region of M~31. The two \hii\ regions observed in M~33, 
NGC~595 and NGC~604, were  previously studied by \cite{vilchezetal88}. \cite{estebanetal02} detected the \cii\ and \oii\ lines for the first time in NGC~604. 
There are no previous detections of those RLs in NGC~595.   
The nucleus --zone C-- of the starburst galaxy NGC~1741 has been studied by \citet{lopez-sanchezetal04}. 
These authors reported a dubious detection of a relatively bright emission feature temptatively identified 
as \cii\ $\lambda$4267 that has not been confirmed in our spectra. The giant \hii\ region 
NGC~2363 in the irregular galaxy NGC~2366 has very detailed previous spectroscopical studies, such as those by \cite{gonzalez-delgadoetal94} and \cite {estebanetal02}. These last authors detected \cii\ and \oii\ lines for the first time in 
NGC~2363. The three \hii\ regions observed in NGC~2403, VS~24, VS~38, and VS~44, were studied by \cite{garnettetal97}. 
There are no previous detections of \cii\ and \oii\ lines for these objects. Region \#70 of the spiral galaxy NGC~4395 had only 
very scarce spectral data available \citep{mccalletal85,cedrescepa02} and no electron temperature determinations. 
Optical spectra of the brightest region of the \hii\ galaxy NGC~4861 were analyzed by \cite{dinersteinshields86}. A very  
deep spectrum of the center of M~83 has been obtained by \cite{bresolinetal05}, who determine electron temperatures  
from more emission line ratios than in this study. It has been surprising not detecting \cii\ and \oii\ lines in 
our HIRES spectrum of the central \hii\ region of M~83 considering the high surface brightness and high metallicity of the nebula. However, 
this is a very low 
ionization degree object, its stellar underlying continuum is very high, and --most importantly-- its emission lines are extremely broad. 
We consider that the high-spectral resolution used in our observation oversample too much the line profiles, producing a dramatical decrease of 
the intensity contrast of the fainter lines with respect to the strong continuum that prevents their detection. 
Finally, the three \hii\ regions of M~101 observed in this paper, have also been studied 
in several works. H~1013 has been analyzed in depth by \cite{bresolin07}, who was the first in detecting \cii\ lines 
in this object. NGC~5461 has been extensively studied in the literature \citep{rayoetal82,kennicuttgarnett96,estebanetal02}. \cite{estebanetal02} detected \cii\ and \oii\ lines in this object for the first time. Optical spectra of NGC~5447 have been analysed by \cite{torres-peimbertetal89} and \cite{kennicuttgarnett96}, but they do not determine electron temperatures, 
while \cite{kennicuttetal03} did so in three sub-components (H128, H143, H149) of this extended giant \hii\ region. 
The line intensities reported in this paper do not show systematic differences with respect to those published in previous papers. 
They are consistent considering the expected differences due to aperture effects and/or extinction coefficients and reddening laws used. 

It is interesting to note the identification of some lines of {\cai}] at $\lambda\lambda$ 7394.08, 7423.6, 7442.99, 7457.3,   
and {\cai} $\lambda$ 7451.87 in the spectra of some of the sample objects. In particular, {\cai} $\lambda$ 7451.87 and {\cai}] $\lambda$ 7457.3 are rather bright, with derredened intensities of the order of 1-3 \% of $I$(H$\beta$), and  
have been observed in 11 out of 13 objects for which we have spectra covering the spectral range redwards 7400 \AA. We have carefully verified that all these 
lines do not correspond to sky emission features and that the wavelength calibration is accurate in that part of the spectrum. 
Therefore, they are real emission lines produced in the \hii\ regions.  
It is striking that {\cai}] and {\cai} lines have not been previously identified in other \hii\ regions from high-resolution echelle spectra published by our group \citep[e.g.][]{garcia-rojas06,lopez-sanchezetal07}. Moreover, we have not found references about their identification in 
other Galactic or extragalactic \hii\ regions in the literature. The only exception is the detection of about 11 {\cai}] lines 
in the spectral range between 7890 and 9700~\AA\ in an unpublished VLT echelle spectrum that we obtained of a bright zone of the 
bar of the Orion Nebula. All the {\cai}] lines detected in this paper are intercombination lines between the spectra terms 
$^3$F$^0$ and $^1$D, and the single {\cai} line detected corresponds to a $^3$F$^0-^3$D transition. Consistently, all the 
{\cai}] lines detected in the bar of the Orion Nebula are also intercombination ones but, in this case, half of them correspond to a $^3$P$^0-^1$D transition. To find an explanation for the presence of these lines in the spectra is outside the scope of the 
present paper. However, the fact that similar lines have only been detected in the bar of the Orion Nebula and not in other 
similar spectra --taken with the same instrument and with the same exposure time-- of the central parts of the nebula 
\citep[see][]{estebanetal04}, 
suggests that the {\cai}] emission should be produced in the ionization edge of the EHRs observed in this paper, perhaps due 
to a fluorescence process at the photodissociation region (PDR). It would be interesting to further investigate the possible 
mechanism originating these transitions. 

The observed line intensities of our objects must be corrected for interstellar reddening. This can be done using the reddening constant, 
$c$(H$\beta$), obtained from the intensities of the Balmer lines. However, the fluxes of {\hi} lines may be also 
affected by underlying stellar absorption. Consequently, we have performed an iterative procedure to derive both 
$c$(H$\beta$) and the equivalent widths of the absorption in the hydrogen lines, $W_{\rm abs}$, which we 
use to correct the observed line intensities. We assumed that the equivalent width of
the absorption components is the same for all the Balmer lines
and used the relation given by \citet{mazzarellaboroson93} for the absorption correction of each Balmer line following 
the procedure outlined by \citet{lopez-sanchezetal06}. We have used the reddening curve of \citet{seaton79} and the observed H$\alpha$/H$\beta$, H$\gamma$/H$\beta$, H$\delta$/H$\beta$, and H$\epsilon$/H$\beta$ line ratios. We have considered the theoretical line ratios expected for case B recombination given by \citet{storeyhummer95} for electron densities of 
100 cm$^{-3}$ and a first estimation of the electron temperature of each object based on the de-reddened line intensity ratios. 
In tables~\ref{lineid1}--\ref{lineid3}, we include the $c$(H$\beta$) and $W_{\rm abs}$ pairs that provide the 
best match between the corrected and the theoretical line ratios. This procedure was not applied to NGC~5447 due to the fact 
that the shorter wavelength range covered in this object only includes three Balmer lines. Therefore, we have assumed that $W_{\rm abs}$ = 0 
in the hydrogen lines for this object. In tables~\ref{lineid1}--\ref{lineid3}, we also include the observed (uncorrected for reddening) 
integrated H$\beta$ flux, $F$(H$\beta$), the equivalent width of this line, $W$(H$\beta$), and the equivalent width of the absorption in the hydrogen lines, $W_{\rm abs}$. 

The spectra of NGC~595, H~1013, and NGC~5461 show broad emission features centered at 4640 and 4686 \AA.  
These two features --which are blended when observed at low spectral resolution but not in the present observations-- 
correspond to the so-called Wolf-Rayet, W$-$R, blue bump 
originating from the stellar winds of these evolved massive stars. The brightest feature corresponds to \heii\ $\lambda$4686, and  
the emission at 4640 \AA\ is the blend of \niii\ $\lambda$4634 and \niii\ $\lambda$4640, which are characteristic of W$-$R stars of the WNL subtype \citep{smithetal96}. In the cases of NGC~595, H~1013, and NGC~5461, those emission features were 
previously reported by  \cite{contimassey81}, \cite{bresolin07}, and \cite{rayoetal82}, respectively. 
In the case of NGC~5447, only a broad \heii\ $\lambda$4686 emission has been detected in our spectra. As far as we know, this is the first 
time that the broad \heii\ $\lambda$4686 emission feature is observed in the object. 

In principle, the ratio between the integrated fluxes of the W$-$R blue bump and H$\beta$ can be used to estimate the 
W$-$R/O number ratio of a given star-forming region, however, our slit positions only cover a fraction of the total extension of the region 
and that ratio can not be properly derived from our observations. In any case, the observed 
integrated fluxes (uncorrected for reddening) of the W$-$R bumps at $\lambda$4640 and $\lambda$4686 are also included in Table~\ref{lineid1}. 

\begin{deluxetable}{lccccccccccccc} 
\tabletypesize{\scriptsize}
\rotate
\tablecaption{Dereddened line intensity ratios with respect to I(H$\beta$)=100 of NGC~595, NGC~604, H~1013, NGC~5461, and NGC~5447.
\label{lineid1}}
\tablewidth{0pt}
\tablehead{
\multicolumn{4}{c}{} & \multicolumn{2}{c}{NGC~595} & \multicolumn{2}{c}{NGC~604} & \multicolumn{2}{c}{H~1013} & \multicolumn{2}{c}{NGC~5461} & \multicolumn{2}{c}{NGC~5447}\\
\cline{5-6} \cline{7-8} \cline{9-10} \cline{11-12} \cline{13-14}
\colhead{$\lambda_0$} & \multicolumn{3}{c}{} & \colhead{$\lambda$} & \colhead{} & \colhead{$\lambda$} & \colhead{} & \colhead{$\lambda$} & \colhead{} 
& \colhead{$\lambda$} & \colhead{} & \colhead{$\lambda$} & \colhead{} \\
\colhead{(\AA)} & \colhead{Ion} & \colhead{ID} & \colhead{$f(\lambda)$}  & \colhead{(\AA)} &\colhead{$I(\lambda)$} & \colhead{(\AA)} &\colhead{$I(\lambda)$} & \colhead{(\AA)} &\colhead{$I(\lambda)$} & \colhead{(\AA)} &\colhead{$I(\lambda)$} & \colhead{(\AA)} &\colhead{$I(\lambda)$}} 
\startdata

3587.28	&	He I	&	32	&	0.278	&	3584.90	&	0.17	$\pm$	0.05	&	\nd	&			\nd	&	\nd	&			\nd	&	3590.18	&	0.4	$\pm$	0.1 &	\nd	&	\nd	\\		
3613.64	&	He I	&	6	&	0.275	&	3611.19	&	0.26	$\pm$	0.05	&	3610.58	&	0.22	$\pm$	0.05	&	\nd	&			\nd	&	3616.79	&	0.33	$\pm$	0.09 &	\nd	&	\nd	\\		
3634.25	&	He I	&	28	&	0.272	&	3631.75	&	0.35	$\pm$	0.05	&	3631.20	&	0.24	$\pm$	0.06	&	3637.90	&	0.29	$\pm$	0.08	&	3637.50	&	0.21	$\pm$	0.08 &	\nd	&	\nd	\\				
3663.4	&	H I	&	H29	&	0.267	&	\nd	&			\nd	&	3660.35	&	0.21	$\pm$	0.05	&	\nd	&			\nd	&	\nd	&			\nd &	\nd	&	\nd	\\		
3664.68	&	H I	&	H28	&	0.267	&	\nd	&			\nd	&	3661.56	&	0.25	$\pm$	0.06	&	\nd	&			\nd	&	\nd	&			\nd &	\nd	&	\nd	\\		
3666.1	&	H I	&	H27	&	0.267	&	3663.63	&	0.19	$\pm$	0.05	&	3662.97	&	0.26	$\pm$	0.06	&	\nd	&			\nd	&	\nd	&			\nd &	\nd	&	\nd	\\		
3667.68	&	H I	&	H26	&	0.266	&	3665.22	&	0.25	$\pm$	0.05	&	3664.54	&	0.23	$\pm$	0.05	&	3671.26	&	0.17:	&	3670.89	&	0.10: &	\nd	&	\nd	\\		
3669.47	&	H I	&	H25	&	0.266	&	3667.00	&	0.29	$\pm$	0.05	&	3666.40	&	0.21	$\pm$	0.05	&	3673.05	&	0.28	$\pm$	0.08	&	3672.64	&	0.13: &	\nd	&	\nd	\\
\enddata
\tablenotetext{a}{Blended with the W$-$R blue bump emission at 4640 \AA.} 
\tablenotetext{b}{Blend of O II $\lambda$4649.13 and $\lambda$4650.84 lines. }
\tablenotetext{c}{Blend of O I $\lambda$6046.23, $\lambda$6046.44, and $\lambda$6046.49 lines. }
\tablenotetext{d}{Blend of [O II] $\lambda$7318.39 and $\lambda$7319.99 lines.}
\tablenotetext{e}{Blend of [O II] $\lambda$7329.66 and $\lambda$7330.73 lines.}
\end{deluxetable}

\begin{deluxetable}{lccccccccccccc} 
\tabletypesize{\scriptsize}
\rotate
\tablecaption{Dereddened line intensity ratios with respect to I(H$\beta$)=100 of VS~24, VS~38, VS~44, NGC~2363, and K932
\label{lineid2}}
\tablewidth{0pt}
\tablehead{
\multicolumn{4}{c}{} & \multicolumn{2}{c}{VS~24} & \multicolumn{2}{c}{VS~38} & \multicolumn{2}{c}{VS~44} & \multicolumn{2}{c}{NGC~2363} & \multicolumn{2}{c}{K~932}\\
\cline{5-6} \cline{7-8} \cline{9-10} \cline{11-12} \cline{13-14}
\colhead{$\lambda_0$} & \multicolumn{3}{c}{} & \colhead{$\lambda$} & \colhead{} & \colhead{$\lambda$} & \colhead{} & \colhead{$\lambda$} & \colhead{} 
& \colhead{$\lambda$} & \colhead{} & \colhead{$\lambda$} & \colhead{} \\
\colhead{(\AA)} & \colhead{Ion} & \colhead{ID} & \colhead{$f(\lambda)$}  & \colhead{(\AA)} &\colhead{$I$($\lambda$)} & \colhead{(\AA)} &\colhead{$I$($\lambda$)} & \colhead{(\AA)} &\colhead{$I$($\lambda$)} & \colhead{(\AA)} &\colhead{$I$($\lambda$)} & \colhead{(\AA)} &\colhead{$I$($\lambda$)}}
\startdata
3554.42	&	He I	&	34	&	0.283	&	\nd	&			\nd	&	\nd	&			\nd	&	\nd	&			\nd	& 3555.37	&	0.22	$\pm$	0.03	&	\nd	&			\nd	\\
3587.28	&	He I	&	32	&	0.278	&	\nd	&			\nd	&	\nd	&			\nd	&	3589.02	&	0.22	$\pm$	0.06	&	3588.20	&	0.17	$\pm$	0.03	&	3586.52	&	0.34	$\pm$	0.08	\\
3613.64	&	He I	&	6	&	0.275	&	3614.41	&	0.3	$\pm$	0.1	&	\nd	&			\nd	&	3615.24	&	0.23	$\pm$	0.06	&	3614.55	&	0.22	$\pm$	0.03	&	3612.91	&	0.46	$\pm$	0.09	\\
3634.25	&	He I	&	28	&	0.272	&	3634.96	&	0.3	$\pm$	0.1	&	\nd	&			\nd	&	3635.86	&	0.12:	&	3635.19	&	0.29	$\pm$	0.03	&	3633.51	&	0.49	$\pm$	0.09	\\
3661.22	&	H I	&	H31	&	0.267	&	\nd	&			\nd	&	\nd	&			\nd	&	\nd	&			\nd	&	3662.20	&	0.05:	&	3660.49	&	0.27	$\pm$	0.08	\\
3662.26	&	H I	&	H30	&	0.267	&	3662.99	&	0.2:	&	\nd	&			\nd	&	\nd	&			\nd	&	3663.18	&	0.07	$\pm$	0.02	&	3661.54	&	0.23	$\pm$	0.08	\\
3663.4	&	H I	&	H29	&	0.267	&	3664.19	&	0.2:	&	\nd	&			\nd	&	\nd	&			\nd	&	3664.30	&	0.12	$\pm$	0.03	&	3662.67	&	0.34	$\pm$	0.08	\\
3664.68	&	H I	&	H28	&	0.267	&	3665.47	&	0.20:	0.09	&	\nd	&			\nd	&	\nd	&			\nd	&	3665.62	&	0.15	$\pm$	0.03	&	3663.84	&	0.21	$\pm$	0.08	\\
\enddata
\tablenotetext{a}{Blend of [O II] $\lambda$7318.39 and $\lambda$7319.99 lines.}
\tablenotetext{b}{Blend of [O II] $\lambda$7329.66 and $\lambda$7330.73 lines.}
\end{deluxetable}

\begin{deluxetable}{lccccccccccc} 
\tabletypesize{\scriptsize}
\rotate
\tablecaption{Dereddened line intensity ratios with respect to I(H$\beta$)=100 of NGC~1741-C, NGC~4395-70, NGC~4861, and M83-Nucleus.
\label{lineid3}}
\tablewidth{0pt}
\tablehead{
\multicolumn{4}{c}{} & \multicolumn{2}{c}{NGC~1741-C} & \multicolumn{2}{c}{NGC~4395-70} & \multicolumn{2}{c}{NGC~4861} & \multicolumn{2}{c}{M83-Nucleus} \\
\cline{5-6} \cline{7-8} \cline{9-10} \cline{11-12}
\colhead{$\lambda_0$} & \multicolumn{3}{c}{} & \colhead{$\lambda$} & \colhead{} & \colhead{$\lambda$} & \colhead{} & \colhead{$\lambda$} & \colhead{} 
& \colhead{$\lambda$} & \colhead{} \\
\colhead{(\AA)} & \colhead{Ion} & \colhead{ID} & \colhead{$f(\lambda)$}  & \colhead{(\AA)} &\colhead{$I$($\lambda$)} & \colhead{(\AA)} &\colhead{$I$($\lambda$)} & \colhead{(\AA)} &\colhead{$I$($\lambda$)} & \colhead{(\AA)} &\colhead{$I$($\lambda$)}}
\startdata
3554.42	&	He I	&	34	&	0.283	&	\nd	&			\nd	&	\nd	&			\nd	&	3564.15	&	0.19	$\pm$	0.05	&	\nd	&			\nd	\\
3587.28	&	He I	&	32	&	0.278	&	\nd	&			\nd	&	\nd	&			\nd	&	3597.06	&	0.22	$\pm$	0.05	&	\nd	&			\nd	\\
3613.64	&	He I	&	6	&	0.275	&	\nd	&			\nd	&	\nd	&			\nd	&	3623.51	&	0.31	$\pm$	0.06	&	\nd	&			\nd	\\
3634.25	&	He I	&	28	&	0.272	&	\nd	&			\nd	&	\nd	&			\nd	&	3644.13	&	0.35	$\pm$	0.06	&	\nd	&			\nd	\\
3666.1	&	H I	&	H27	&	0.267	&	\nd	&			\nd	&	\nd	&			\nd	&	3675.99	&	0.18	$\pm$	0.05	&	\nd	&			\nd	\\
3667.68	&	H I	&	H26	&	0.266	&	\nd	&			\nd	&	\nd	&			\nd	&	3677.76	&	0.22	$\pm$	0.05	&	\nd	&			\nd	\\
3669.47	&	H I	&	H25	&	0.266	&	\nd	&			\nd	&	\nd	&			\nd	&	3679.47	&	0.26	$\pm$	0.05	&	\nd	&			\nd	\\
3671.48	&	H I	&	H24	&	0.266	&	\nd	&			\nd	&	\nd	&			\nd	&	3681.45	&	0.34	$\pm$	0.06	&	\nd	&			\nd	\\
\enddata
\tablenotetext{a}{Blend of [O II] $\lambda$7318.39 and $\lambda$7319.99 lines.}
\tablenotetext{b}{Blend of [O II] $\lambda$7329.66 and $\lambda$7330.73 lines.}
\end{deluxetable}

\begin{deluxetable}{lcccccccc} 
\tabletypesize{\scriptsize}
\rotate
\tablecaption{Physical conditions
\label{physcond}}
\tablewidth{0pt}
\tablehead{
\colhead{Parameter}  & 
\colhead{Lines}  & 
\colhead{NGC~595} & \colhead{NGC~604} & \colhead{H1013} & \colhead{NGC~5461} & \colhead{NGC~5447} & \colhead{VS~24} & \colhead{VS~38}}   
\startdata 
$N_{\rm e}$ (cm$^{-3}$) & [N I] & 270$\pm$180 & 140$\pm$100 & 114$^{+140}_{-114}$ & $<$100 & \nd & $<$100 & 220$^{+320}_{-220}$ \\
& [O II] & 260$\pm$30 & 270$\pm$30 & 280$\pm$60 & 540$\pm$110 & \nd & 450$\pm$40 & 310$\pm$30  \\
& [S II] & $<$100 & $<$100 & $<$100 & 260$\pm$160 & 220$\pm$170 & 120$\pm$50 & $<$100  \\
& [Cl III] & 700$\pm$370 & 490$^{+510}_{-490}$ & 70$^{+780}_{-70}$ & 180$^{+460}_{-180}$ & 280$^{+690}_{-280}$ & 640$^{+930}_{-640}$ & 50$^{+1800}_{-50}$ \\
& [Ar IV] & \nd &  \nd & \nd & \nd & 1170$^{+2020}_{-1170}$ &\nd & \nd \\
$T_{\rm e}$ (K) & [N II] & 8400$\pm$160 & 8950$\pm$270 & 7970$\pm$370 & 9310$\pm$400 & 10950$\pm$610 &8530$\pm$340 & 8260$\pm$470 \\
& [O II] & 7760$\pm$100 & 9190$\pm$160 & 6010$\pm$370 & \nd & \nd & 7810$\pm$160 & 7720$\pm$150 \\
& [S II] & 7070$\pm$260 & 8350$\pm$330 & \nd & 7280$\pm$500 & \nd & 6990$\pm$420 & \nd \\
\enddata
\end{deluxetable}

\section{Physical Conditions and Chemical Abundances}
\label{phiscondabund} 
\subsection{Electron Temperatures and Densities}
\label{temden}
The physical conditions of the ionized gas: electron temperatures, {\elect}, and densities, {\elecd}, have been derived 
from the usual CEL ratios, using the 
{\sc IRAF} task {\tt temden} of the package NEBULAR \citep{shawdufour95} with updated atomic data for some ions \citep[see][their table 4]{garcia-rojasetal05}. Electron densities have been derived from [{\nitroi}] $\lambda\lambda$5198/5200, [{\oii}] $\lambda\lambda$3726/3729, [{\sii}] $\lambda\lambda$6717/6731, [{\cliii}] $\lambda\lambda$5518/5538, and [{\ariv}] $\lambda\lambda$4711/4740  
line ratios. Electron temperatures have been calculated using four sets of auroral to nebular line intensity ratios: [{\nii}] $\lambda$5755/[{\nii}] $\lambda\lambda$6548, 6583, [{\oii}] $\lambda\lambda$7319, 7330/[{\oii}] $\lambda\lambda$3726, 3729, [{\sii}]$\lambda\lambda$4069, 4076/[{\sii}] $\lambda\lambda$6717, 6731, and [{\oiii}] $\lambda$4363/[{\oiii}] $\lambda\lambda$4959, 5007.  
The procedure for the determination of the physical conditions was the following: an initial {\elect}-value of 10,000 K
was assumed in order to derive a first approximation to the different {\elecd} determinations; then these preliminary {\elecd}-values were used to recompute {\elect}, and finally, we iterated until convergence to obtain the adopted values 
of {\elecd} and {\elect}. {\elect}([{\oii}]) and {\elect}([{\nii}]) have been corrected from the contribution to 
[{\oii}] $\lambda\lambda$7319, 7330 and [{\nii}] $\lambda$5755, respectively, due to recombination following the formulae
derived by \citet{liuetal00}. These contributions are between 4 and 6\% in the case of the intensity of [{\oii}] $\lambda\lambda$7319, 7330 lines and between 1 and 3\% for the intensity of [{\nii}] $\lambda$5755. It is necessary to estimate the N$^{++}$/H$^+$ ratio for determining the recombination contribution to the intensity of the auroral line of 
[{\nii}]. In all the cases, we have estimated this abundance ratio considering a preliminary determination of the N$^{+}$, O$^{+}$, and O$^{++}$ abundances and assuming that O$^{++}$/O$^+$ is approximately N$^{++}$/N$^+$. In the case of NGC~2363, its {\elect}([{\oii}]) is above the range of validity (5000 to 10,000 K) of the formula of \cite{liuetal00} for the correction due to recombination of the [{\oii}] $\lambda\lambda$7319, 7330 lines. Therefore, the correction we estimate for this object could not be correct. However, the fact that the recombination rate decreases with temperature implies that this effect should be very small in NGC~2363. 
The physical conditions determined for all the sample objects are shown 
in Table~\ref{physcond}. 

All the sample objects show low densities, with values lower that 10$^3$ cm$^{-3}$ and even near the low density limit in some cases. The central {\hii} region of M~83 shows the largest {\elecd}, three different indicators giving consistent results of about 700 cm$^{-3}$. There are three objects where the {\elecd} has been derived from the [{\ariv}] line ratio, in all the cases this indicator gives larger densities than the other line ratios and about 10$^3$ cm$^{-3}$, suggesting some density stratification in these nebulae. In fact, Ar$^{3+}$ is the ion with the highest ionization potential for which 
we can derive {\elecd}, and its associated Str\"omgren sphere should be located in the innermost parts of the nebula. 

\begin{figure*}
  \includegraphics[angle=0,scale=1]{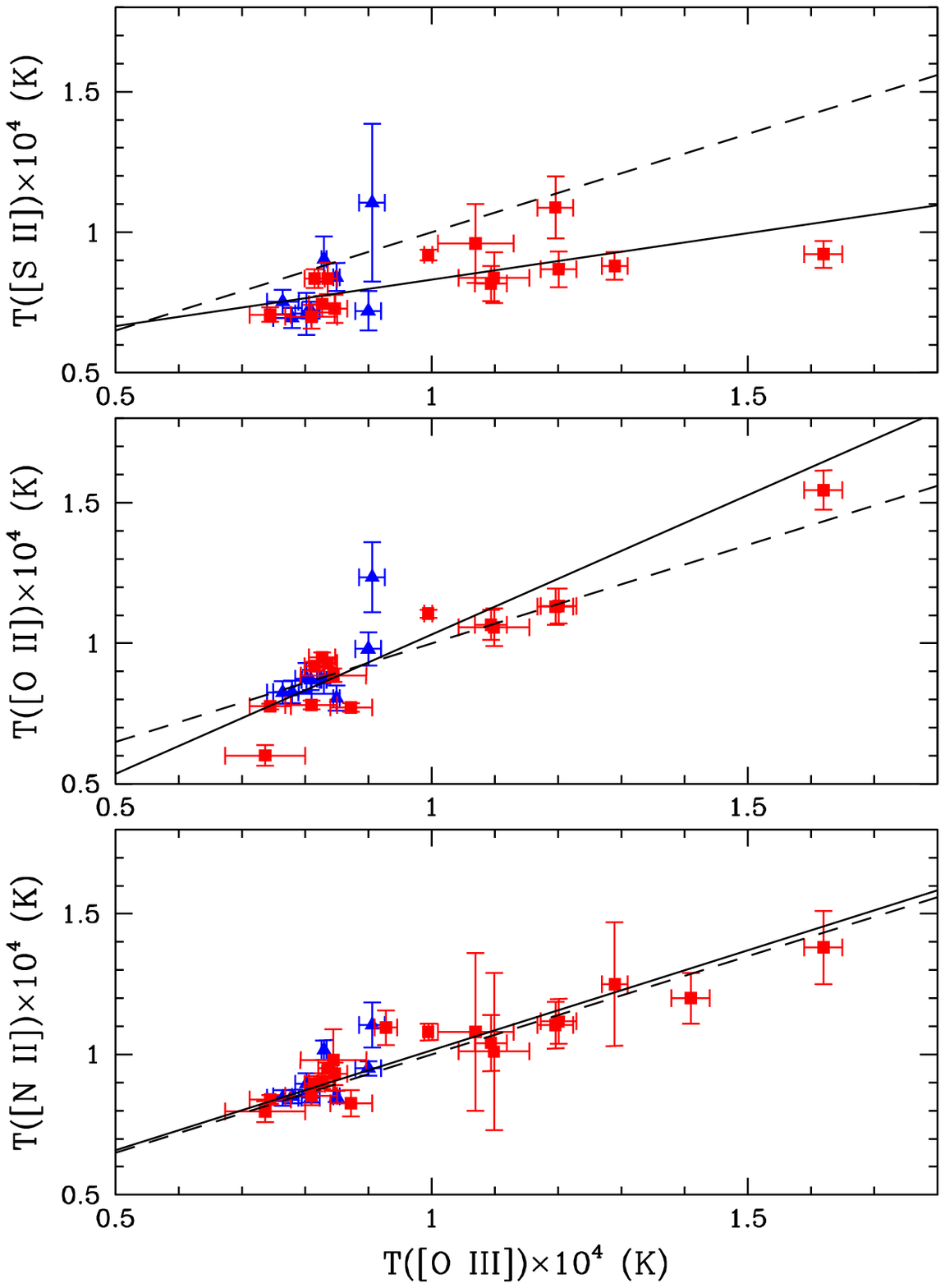} 
  \caption{Correlations between various {\elect} diagnostics. Filled (red) squares show our measurements for extragalactic 
  \hii\ regions and those of NGC~5253 from \citet{lopez-sanchezetal07}, 30 Doradus from \citet{peimbert03}, and NGC~5471 from 
  \citet{estebanetal02}. Filled (blue) triangles show measurements for a sample of Galactic \hii\ regions taken from the compilation by  
  \citet{garcia-rojasesteban07}. The solid lines represent the least-squares fit of the measurements. The dashed lines 
  show the predicted correlations from \citet{garnett92} based on photoionization models.}
  \label{T_relations}
\end{figure*}

The data compiled in Table~\ref{physcond} provides a relatively large number of different {\elect} determinations, in particular those associated with low ionization potential ions: [{\nii}], [{\oii}], and [{\sii}]. This allows us to compare the consistency of different temperature scales. 
\citet{garnett92} constructed photoionization models that provide simple scaling relations between temperatures measured from different ions. This author obtains the following relation between {\elect}([{\oiii}]) and other temperature indicators of low ionization degree ions --such as {\elect}([{\nii}]), {\elect}([{\oii}]), 
and {\elect}([{\sii}])-- valid for the temperature range from 2000 to 18,000~K: 
\begin{eqnarray}
\label{garnett}
 T_e([{\rm N\medspace II}]) = T_e([{\rm O\medspace II}]) = T_e([{\rm S\medspace II}]) = 0.70 \times T_e([{\rm O\medspace III}]) + 3000~ {\rm K} .
\end{eqnarray}
Similar relationships between {\elect}([{\oiii}]) and {\elect}([{\oii}]) have been obtained by other authors  \citep{campbelletal86,izotovetal94,pilyuginetal06} and have been extensively used in many works. Several authors have tried to make observational tests of the reliability of these relations  \citep{garnettetal97,kennicuttetal03,perez-monterodiaz03,hageleetal06,hageleetal08,bresolinetal09}. 
The {\elect}([{\nii}]) vs. {\elect}([{\oiii}]) relationship has been the most difficult one to test because of the paucity of 
good determinations of the {\elect}([{\nii}]) temperature indicator. This is because most studies have focused on the determination of 
chemical abundances in low-metallicity, high-excitation EHRs, especially \hii\ galaxies, where the [{\nii}] $\lambda$ 
5755 line is very faint \citep{perez-monterodiaz03,hageleetal06,hageleetal08}. \citet{kennicuttetal03} 
studied \hii\ regions in M~101, a group of objects whose properties are more similar to those of most of our EHRs, but the small number of objects with {\elect}([{\nii}]) determinations in their sample does not permit to obtain a clear trend between {\elect}([{\nii}]) and {\elect}([{\oiii}]), indicating that more data were needed to draw meaningful conclusions. 
In Figure~\ref{T_relations}, we 
show the relations between {\elect}([{\nii}]), {\elect}([{\oii}]), and {\elect}([{\sii}]) with respect to {\elect}([{\oiii}]) obtained from the data presented in this paper and those of NGC~5253 \citep{lopez-sanchezetal07}, 30 Doradus \citep{peimbert03}, NGC~5471 \citep{estebanetal02}, and the compilation of data for a sample of Galactic \hii\ regions of \citet{garcia-rojasesteban07}. These data have been included because they are also based on high-spatial resolution spectroscopy 
(except in the cases of NGC~5447 and NGC~5471, where intermediate-resolution spectroscopy was used) and they have been analyzed in the same manner as the present sample, using the same atomic dataset and considering the recombination contribution to both the [{\nii}] and [{\oii}] line intensities. In the figure, we can see a rather clear linear relation between {\elect}([{\nii}]) with respect to {\elect}([{\oiii}]), which is a susbtantial improvement with respect to the previous results by \citet{kennicuttetal03}. Our least-squares fit to the points --weighted by their observational errors-- of the {\elect}([{\nii}]) vs. {\elect}([{\oiii}]) diagram gives:  
\begin{eqnarray}
\label{n2vso3} 
 T_e([{\rm N\medspace  II}]) = 0.71 \times T_e([{\rm O\medspace III}]) + 3050~ {\rm K} ,
\end{eqnarray}
with a correlation coefficient of 0.91. It is remarkable that this empirical relation is almost identical to the theoretically predicted relationship from \citet{garnett92}, confirming that the use of Garnett's relation between these two indicators 
is entirely reliable (the same conclusion has been drawn by \citealt{bresolinetal09}). This is a result of special interest because while {\elect}([{\oiii}]) usually cannot be derived in metal-rich \hii\ regions --where the 
[{\oiii}] $\lambda$4363 auroral line becomes very faint-- {\elect}([{\nii}]) can be more easily determined from optical spectra 
\citep[see][]{bresolin06}, and shows a very weak dependence on {\elecd}.  

The apparent lack of consistency between {\elect}([{\oii}]) and {\elect}([{\oiii}]) found by all the authors that have studied the empirical relation of these temperature indicators \citep{kennicuttetal03,perez-monterodiaz03,hageleetal06,hageleetal08} is remarkable. \citet{kennicuttetal03} proposed several possible sources of such inconsistency: 
a) the recombination contribution to [{\oii}] line intensity, b) the large  contribution of collisional de-excitation to the [{\oii}] line ratios --dependence on {\elecd}-- used for deriving {\elect}([{\oii}]), which is more 
important than for the rest of the indicators, except {\elect}([{\sii}]), c) radiative transfer effects, d) observational uncertainties in the interstellar reddening, e) contamination by OH airglow emission of the [{\oii}] multiplet around 7325 \AA. 
\citet{perez-monterodiaz03} and \citet{hageleetal06,hageleetal08} obtain also a large scatter in their {\elect}([{\oii}]) vs. {\elect}([{\oiii}]) diagrams, that include a large number of data points from 
the literature. These authors indicate that the scatter is 
probably due to the dependence of {\elect}([{\oii}]) on electron density. However, our results shown in Figure~\ref{T_relations} indicate a tighter and rather linear relation between {\elect}([{\oii}]) and {\elect}([{\oiii}]). The least-squares fit gives: 
\begin{eqnarray}
 T_e([{\rm O\medspace II}]) = 0.99 \times T_e([{\rm O\medspace III}]) + 410~ {\rm K} ,
\end{eqnarray}
with a correlation coefficient of 0.88, indicating that the use of Garnett's relationship 
between {\elect}([{\oii}]) and {\elect}([{\oiii}]) seems to be a reasonable approximation. It is important to remark that our data 
have been corrected for the recombination contribution to [{\oii}] --as also done by \citet{perez-monterodiaz03,hageleetal06,hageleetal08}-- and that the use of high spectral-resolution 
spectra precludes the contamination by OH airglow emission of the red [{\oii}] lines. The comparison of our result with the previous ones from the literature is puzzling. 
Considering that most of the scatter shown in the {\elect}([{\oii}]) vs. {\elect}([{\oiii}]) diagrams of 
\citet{perez-monterodiaz03} and \citet{hageleetal06,hageleetal08} should be due to the inclusion of inhomogeneous data points from the literature, we 
think that the sensitivity of {\elect}([{\oii}]) to many different factors makes  the statistical use of this indicator ill-advised
when using inhomogeneous datasets. 

As it can be seen in Figure~\ref{T_relations}, the third relationship studied: {\elect}([{\sii}]) vs. {\elect}([{\oiii}]) shows a somewhat different behaviour than the previous ones. The data points show a larger scatter and the least-squares fit gives: 
\begin{eqnarray}
 T_e([{\rm S\medspace II}]) = 0.33 \times T_e([{\rm O\medspace III}]) + 5015~ {\rm K} ,
\end{eqnarray}
with a correlation coefficient of 0.53, lower than in the previous cases. A large scatter in this relation is also evident 
in the data obtained by \citet{perez-monterodiaz03} and \citet{hageleetal08}. A common result of these works is that  
{\elect}([{\sii}]) is smaller than {\elect}([{\oii}]) in most cases, in agreement with our results. There are several arguments to explain this behavior: a) due to the lower ionization potential of neutral sulphur, the zone of S$^+$ should not coincide with the zone where N$^+$ and O$^+$ are present --which are essentially cospatial-- and this may be reflected in somewhat different temperatures in both zones, b) the contribution of [{\sii}] emission from diffuse ionized gas due to photon leakage from the \hii\ region, c) the 
dependence of {\elect}([{\sii}]) on electron density, d) some recombination contribution to [{\sii}] line intensities. 

\begin{deluxetable}{lcccccccccc} 
\tabletypesize{\scriptsize}
\rotate
\tablecaption{Ionic and Total Abundances{\TA}
\label{abund1}}
\tablewidth{0pt}
\tablehead{
& \multicolumn{2}{c}{NGC~595} & \multicolumn{2}{c}{NGC~604} &  \multicolumn{2}{c}{H1013} & \multicolumn{2}{c}{NGC~5461}  & \multicolumn{2}{c}{NGC~5447}\\
\cline{2-3} \cline{4-5} \cline{6-7} \cline{8-9} \cline{10-11}
 & 
\colhead{$t^2$ = 0.000} & \colhead{$t^2$ = 0.036} & \colhead{$t^2$ = 0.000} & \colhead{$t^2$ = 0.034} & 
\colhead{$t^2$ = 0.000} & \colhead{$t^2$ = 0.037} & \colhead{$t^2$ = 0.000} & \colhead{$t^2$ = 0.014} & \colhead{$t^2$ = 0.000} & \colhead{$t^2$ = 0.032}}
\startdata 
\multicolumn{11}{c}{Ionic Abundances from Collisionally Excited Lines} \\
\hline
N$^+$ & 7.13$\pm$0.02 & 7.28$\pm$0.06 & 6.82$\pm$0.03 & 6.94$\pm$0.04 & 7.35$\pm$0.04 & 7.52$\pm$0.11 & 6.85$\pm$0.04 & 6.89$\pm$0.06 
& 6.43$\pm$0.05 & 6.51$\pm$0.05\\
O$^+$ & 8.27$\pm$0.03 & 8.44$\pm$0.06 & 7.92$\pm$0.04 & 8.06$\pm$0.05 & 8.23$\pm$0.06 & 8.43$\pm$0.13 & 7.80$\pm$0.06 & 7.85$\pm$0.08 
& 7.57$\pm$0.11 & 7.66$\pm$0.11 \\
O$^{++}$ & 7.99$\pm$0.06 & 8.33$\pm$0.07 & 8.19$\pm$0.03 & 8.45$\pm$0.04 & 8.06$\pm$0.12 & 8.42$\pm$0.12 & 8.29$\pm$0.03 & 8.38$\pm$0.07 & 
8.27$\pm$0.02 & 8.46$\pm$0.06 \\
Ne$^{++}$ & 6.73$\pm$0.08 & 7.11$\pm$0.16 & 7.27$\pm$0.03 & 7.56$\pm$0.08 & 7.01$\pm$0.16 & 7.41$\pm$0.36 & 7.46$\pm$0.04 & 7.56$\pm$0.11 & 
\nd & \nd \\ 
S$^+$ & 5.50$\pm$0.02 & 5.65$\pm$0.06 & 5.47$\pm$0.03 & 5.59$\pm$0.04 & 5.63$\pm$0.04 & 5.80$\pm$0.11 & 5.34$\pm$0.04 & 5.38$\pm$0.06 & 
5.23$\pm$0.05 & 5.31$\pm$0.05 \\ 
S$^{++}$ & \nd & \nd & \nd & \nd & 6.88$\pm$0.23 & 7.29$\pm$0.40 & 6.76$\pm$0.06 & 6.86$\pm$0.12 & 6.58$\pm$0.06 & 6.79$\pm$0.12 \\
Cl$^{++}$ & 4.80$\pm$0.06 & 5.12$\pm$0.13 & 4.58$\pm$0.03 & 4.83$\pm$0.07 & 4.72$\pm$0.12 & 5.07$\pm$0.27 & 4.59$\pm$0.04 & 4.68$\pm$0.10 & 
4.55$\pm$0.04 & 4.73$\pm$0.09\\ 
Ar$^{++}$ & 6.14$\pm$0.06 & 6.42$\pm$0.12 & 6.04$\pm$0.03 & 6.26$\pm$0.06 & 6.15$\pm$0.12 & 6.45$\pm$0.23 & 6.09$\pm$0.05 & 6.16$\pm$0.09 & 
6.02$\pm$0.04 & 6.18$\pm$0.08 \\ 
Ar$^{3+}$ & \nd & \nd & \nd & \nd & \nd & \nd & 4.20$\pm$0.10 & 4.29$\pm$0.14 & 4.26$\pm$0.05 & 4.45$\pm$0.10 \\
Fe$^{++}$ & 5.18$\pm$0.17 & 5.34$\pm$0.18 & 5.34$\pm$0.09 & 5.48$\pm$0.09 & 5.55$\pm$0.18 & 5.73$\pm$0.19 & 5.55$\pm$0.18 & 5.60$\pm$0.18 & 
5.38$\pm$0.29 & 5.47$\pm$0.30 \\
\hline
\multicolumn{11}{c}{Ionic Abundances from Recombination Lines} \\
\hline
He$^+$ & 10.834$\pm$0.004 & 10.830$\pm$0.04 & 10.869$\pm$0.005 & 10.866$\pm$0.005 & 10.86$\pm$0.01 & 10.86$\pm$0.01 & 10.86$\pm$0.01 & 10.86$\pm$0.01 & 10.90$\pm$0.02 & 10.88$\pm$0.02 \\
He$^{++}$ & \multicolumn{2}{c}{\nd} & \multicolumn{2}{c}{\nd} & \multicolumn{2}{c}{\nd} & \multicolumn{2}{c}{7.80$\pm$0.11} & \multicolumn{2}{c}{\nd}\\ 
C$^{++}$ & \multicolumn{2}{c}{8.07$\pm$0.12} & \multicolumn{2}{c}{8.19$\pm$0.11} & \multicolumn{2}{c}{8.40$\pm$0.12} & \multicolumn{2}{c}{8.13$\pm$0.20} & \multicolumn{2}{c}{8.06$\pm$0.11}\\ 
O$^{++}$ & \multicolumn{2}{c}{8.33$\pm$0.07} & \multicolumn{2}{c}{8.45$\pm$0.04} & \multicolumn{2}{c}{8.42$\pm$0.11} & \multicolumn{2}{c}{8.38$\pm$0.07} & \multicolumn{2}{c}{8.46$\pm$0.07}\\
\hline
\multicolumn{11}{c}{Total Abundances} \\
\hline
He & 10.849$\pm$0.004 & 10.845$\pm$0.004 & 10.887$\pm$0.005 & 10.884$\pm$0.005 & 10.87$\pm$0.01 & 10.87$\pm$0.01 & 10.88$\pm$0.01 & 10.88$\pm$0.01 & 
10.90$\pm$0.02 & 10.88$\pm$0.02 \\
C & \multicolumn{2}{c}{8.53$\pm$0.12} & \multicolumn{2}{c}{8.40$\pm$0.11} & \multicolumn{2}{c}{8.67$\pm$0.12} & \multicolumn{2}{c}{8.30$\pm$0.20} & \multicolumn{2}{c}{8.20$\pm$0.12} \\ 
N & 7.31$\pm$0.05 & 7.53$\pm$0.10 & 7.28$\pm$0.05 & 7.49$\pm$0.07 & 7.57$\pm$0.10 & 7.82$\pm$0.19 & 7.46$\pm$0.08 & 7.54$\pm$0.11 & 7.21$\pm$0.12 & 7.38$\pm$0.13 \\
O & 8.45$\pm$0.03 & 8.69$\pm$0.05 & 8.38$\pm$0.02 & 8.60$\pm$0.03 & 8.45$\pm$0.06 & 8.73$\pm$0.09 & 8.41$\pm$0.03 & 8.49$\pm$0.06 & 8.35$\pm$0.02 & 8.52$\pm$0.06 \\
Ne & 7.19$\pm$0.10 & 7.47$\pm$0.19 & 7.46$\pm$0.05 & 7.70$\pm$0.10 & 7.41$\pm$0.21 & 7.72$\pm$0.42 & 7.59$\pm$0.06 & 7.67$\pm$0.15 & \nd & \nd \\ 
S & \nd & \nd & \nd & \nd & 6.91$\pm$0.23 & 7.31$\pm$0.48 & 6.86$\pm$0.06 & 6.93$\pm$0.12 & 6.72$\pm$0.06 & 6.88$\pm$0.12 \\
Ar & 6.38$\pm$0.08 & 6.60$\pm$0.13 & 6.22$\pm$0.03 & 6.47$\pm$0.07 & 6.35$\pm$0.14 & 6.62$\pm$0.24 & 6.12$\pm$0.05 & 6.19$\pm$0.09 & 6.04$\pm$0.04 & 6.20$\pm$0.08 \\ 
Fe & 5.34$\pm$0.19 & 5.55$\pm$0.21 & 5.73$\pm$0.10 & 5.94$\pm$0.11 & 5.74$\pm$0.23 & 5.98$\pm$0.26 & 6.08$\pm$0.20 & 6.15$\pm$0.21 & 6.06$\pm$0.33 & 6.23$\pm$0.35 \\
\enddata
\tablenotetext{a}{In units of 12+log(X$^{\rm +m}$/H$^+$).}
\end{deluxetable}

\begin{deluxetable}{lcccccccc} 
\tabletypesize{\scriptsize}
\rotate
\tablecaption{Ionic and Total Abundances{\TA}
\label{abund2}}
\tablewidth{0pt}
\tablehead{
& \colhead{VS~24} & \colhead{VS~38} & \multicolumn{2}{c}{VS~44} & \multicolumn{2}{c}{NGC~2363} & \multicolumn{2}{c}{K932} \\
\cline{2-2} \cline{3-3} \cline{4-5} \cline{6-7} \cline{8-9}
& 
\colhead{$t^2$ = 0.000} & \colhead{$t^2$ = 0.000} & \colhead{$t^2$ = 0.000} & \colhead{$t^2$ = 0.039} & 
\colhead{$t^2$ = 0.000} & \colhead{$t^2$ = 0.120} & \colhead{$t^2$ = 0.000} & \colhead{$t^2$ = 0.033}} 
\startdata 
\multicolumn{9}{c}{Ionic Abundances from Collisionally Excited Lines} \\
\hline
N$^+$ & 7.07$\pm$0.04 & 6.97$\pm$0.05 & 6.90$\pm$0.03 & 7.04$\pm$0.05 & 5.20$\pm$0.06 & 5.43$\pm$0.07 & 6.85$\pm$0.02 & 6.96$\pm$0.03 \\
O$^+$ & 8.09$\pm$0.05 & 8.03$\pm$0.08 & 7.97$\pm$0.04 & 8.13$\pm$0.05 & 6.53$\pm$0.09 & 6.77$\pm$0.05 & 7.83$\pm$0.03 & 7.95$\pm$0.04 \\
O$^{++}$ & 8.05$\pm$0.06 & 7.93$\pm$0.04 & 8.13$\pm$0.03 & 8.43$\pm$0.06 & 7.73$\pm$0.02 & 8.02$\pm$0.05 & 8.28$\pm$0.02 & 8.52$\pm$0.04 \\
Ne$^{++}$ & 6.93$\pm$0.09 & 6.93$\pm$0.06 & 7.16$\pm$0.04 & 7.49$\pm$0.11 & 6.92$\pm$0.02 & 7.23$\pm$0.07 & 7.51$\pm$0.03 & 7.77$\pm$0.07 \\ 
S$^+$ & 5.52$\pm$0.04 & 5.55$\pm$0.05 & 5.49$\pm$0.03 & 5.63$\pm$0.05 & 4.34$\pm$0.06 & 4.56$\pm$0.07 & 5.28$\pm$0.02 & 5.39$\pm$0.03 \\ 
S$^{++}$ & 6.73$\pm$0.12 & 6.37$\pm$0.09 & 6.73$\pm$0.06 & 7.07$\pm$0.11 & 5.70$\pm$0.03 & 6.01$\pm$0.08 & 6.76$\pm$0.04 & 7.03$\pm$0.07 \\
Cl$^{++}$ & 4.58$\pm$0.07 & 4.25$\pm$0.07 & 4.54$\pm$0.04 & 4.83$\pm$0.09 & 3.59$\pm$0.02 & 3.87$\pm$0.07 & 4.61$\pm$0.03 & 4.84$\pm$0.06 \\ 
Cl$^{3+}$ & \nd & \nd & \nd & \nd & 3.84$\pm$0.18 & 4.12$\pm$0.14 & \nd & \nd \\
Ar$^{++}$ & 5.97$\pm$0.06 & 5.88$\pm$0.05 & 6.00$\pm$0.04 & 6.25$\pm$0.08 & 5.19$\pm$0.03 & 5.44$\pm$0.06 & 6.10$\pm$0.03 & 6.30$\pm$0.05 \\ 
Ar$^{3+}$ & \nd & \nd & \nd & \nd & 4.73$\pm$0.02 & 5.03$\pm$0.07 & 3.89$\pm$0.08 & 4.14$\pm$0.08 \\
Fe$^{++}$ & 5.47$\pm$0.17 & 5.36$\pm$0.31 & 5.56$\pm$0.09 & 5.72$\pm$0.09 & 4.36$\pm$0.23 & 4.58$\pm$0.21 & 5.21$\pm$0.07 & 5.32$\pm$0.07 \\
Fe$^{3+}$ & \nd & \nd & \nd & \nd & 5.02$\pm$0.21 & 5.23$\pm$0.22 & \nd & \nd \\
\hline
\multicolumn{9}{c}{Ionic Abundances from Recombination Lines} \\
\hline
He$^+$ & 10.820$\pm$0.007 & 10.881$\pm$0.006 & 10.848$\pm$0.006 & 10.833$\pm$0.006 & 10.88$\pm$0.01 & 10.85$\pm$0.01 & 10.911$\pm$0.005 & 10.907$\pm$0.005 \\
He$^{++}$ & \nd & \nd & \multicolumn{2}{c}{\nd} & \multicolumn{2}{c}{8.34$\pm$0.02} & \multicolumn{2}{c}{\nd} \\ 
C$^{++}$ & 8.15$\pm$0.24 & 8.27$\pm$0.15 & \multicolumn{2}{c}{7.98$\pm$0.18} & \multicolumn{2}{c}{7.55$\pm$0.09} & \multicolumn{2}{c}{8.31$\pm$0.13} \\ 
O$^{++}$ & $<$8.50 & $<$8.54 & \multicolumn{2}{c}{8.43$\pm$0.06} & \multicolumn{2}{c}{8.02$\pm$0.05} & \multicolumn{2}{c}{8.52$\pm$0.04} \\
\hline
\multicolumn{9}{c}{Total Abundances} \\
\hline
He & 10.841$\pm$0.007 & 10.902$\pm$0.006 & 10.863$\pm$0.006 & 10.848$\pm$0.006 & 10.88$\pm$0.01 & 10.85$\pm$0.01 & 10.920$\pm$0.005 & 10.916$\pm$0.005 \\
C & 8.46$\pm$0.24 & 8.50$\pm$0.15 & \multicolumn{2}{c}{8.32$\pm$0.18} & \multicolumn{2}{c}{7.75$\pm$0.09} & \multicolumn{2}{c}{8.46$\pm$0.13} \\ 
N & 7.35$\pm$0.07 & 7.23$\pm$0.11 & 7.29$\pm$0.05 & 7.52$\pm$0.08 & 6.43$\pm$0.07 & 6.70$\pm$0.10 & 7.44$\pm$0.04 & 7.63$\pm$0.06 \\
O & 8.37$\pm$0.04 & 8.29$\pm$0.05 & 8.36$\pm$0.02 & 8.61$\pm$0.04 & 7.76$\pm$0.02 & 8.04$\pm$0.05 & 8.41$\pm$0.02 & 8.62$\pm$0.03 \\
Ne & 7.25$\pm$0.11 & 7.29$\pm$0.09 & 7.39$\pm$0.06 & 7.67$\pm$0.13 & 6.95$\pm$0.03 & 7.25$\pm$0.10 & 7.64$\pm$0.04 & 7.88$\pm$0.09 \\ 
S & 6.77$\pm$0.11 & 6.45$\pm$0.08 & 6.79$\pm$0.06 & 7.12$\pm$0.11 & 5.98$\pm$0.03 & 6.28$\pm$0.07 & 6.85$\pm$0.04 & 7.12$\pm$0.07 \\
Ar & 6.14$\pm$0.06 & 6.06$\pm$0.07 & 6.16$\pm$0.04 & 6.44$\pm$0.08 & 5.32$\pm$0.02 & 5.58$\pm$0.05 & 6.13$\pm$0.04 & 6.32$\pm$0.05 \\ 
Fe & 5.71$\pm$0.19 & 5.58$\pm$0.35 & 5.89$\pm$0.10 & 6.13$\pm$0.12 & 5.44$\pm$0.24 & 5.71$\pm$0.23 & 5.71$\pm$0.08 & 5.90$\pm$0.09 \\
\enddata
\tablenotetext{a}{In units of 12+log(X$^{\rm +m}$/H$^+$).}
\end{deluxetable}

\subsection{Ionic Abundances from CELs and RLs}
\label{ionabund} 
Ionic abundances of N$^+$, O$^+$, O$^{++}$, Ne$^{++}$, S$^+$, S$^{++}$, Cl$^{++}$, Cl$^{3+}$, Ar$^{++}$, and Ar$^{3+}$ have been 
derived from CELs by making use of the {\sc IRAF} task {\tt ionic} of the package NEBULAR. We have assumed a 
two-zone scheme adopting {\elect}([{\nii}]) for ions with a 
low ionization potential (N$^+$, O$^+$, and S$^+$) and {\elect}([{\oiii}]) for ions with a high ionization potential (O$^{++}$, Ne$^{++}$, S$^{++}$ Cl$^{++}$, Cl$^{3+}$, Ar$^{++}$, and Ar$^{3+}$). All the ionic abundances are included in Tables~\ref{abund1}--\ref{abund3}. We have used neither {\elect}([{\oii}]) nor {\elect}([{\sii}]) 
for the abundance determinations. As  discussed in \S\S~\ref{temden}, the {\elect}([{\oii}]) may be affected by 
several sources of uncertainty and, in principle, may be a less confident temperature indicator. On the other hand,
as we pointed out in \S\S~\ref{temden}, {\elect}([{\sii}]) is somewhat lower than {\elect}([{\nii}]) or {\elect}([{\oii}]) 
in most of the objects. This effect has been reported previously in several Galactic \hii\ regions \citep{garcia-rojasetal05,garcia-rojasetal06,garcia-rojasetal07}, and might be produced by the presence of a temperature stratification in the outer zones of the nebulae or, conversely, by uncertainties in the atomic parameters of the ion. 
The \hii\ region at the center of M~83 is the only object without a direct determination of {\elect}([{\oiii}]). For the 
abundance determinations adopted for this nebula we have assumed {\elect}([{\oiii}]) = {\elect}([{\nii}]). The use of 
either the theoretical or the empirical relationships between {\elect}([{\oiii}]) and {\elect}([{\nii}]) shown in  equations~\ref{garnett} or \ref{n2vso3} provides a value of {\elect}([{\oiii}]) of about 2900 K, which gives unreasonably high 
abundances, for example 12+log(O$^{++}$/H$^+$) = 9.72 or even 12+log(Ne$^{++}$/H$^+$) = 10.22. As it has been  
demostrated by \cite{stasinska78,stasinska05}, high-metallicity --and low-temperature-- \hii\ regions may have large temperature 
gradients that could produce differences between the electron temperatures determined from line intensity ratios and the mean ionic temperatures \citep[those obtained from photoionization models and used to derive the equations obtained by][]{garnett92}. This effect can naturally explain the inappropriateness of the 
{\elect}([{\oiii}]) estimated for the \hii\ region at the center of M~83. 

The errors in the ionic abundances shown in tables~\ref{abund1}--\ref{abund3} have been computed as the 
quadratic sum of the independent contributions of {\elecd}, {\elect}, and line flux 
uncertainties. We have considered two sets of abundances, one for {\ts} = 0 and one for {\ts} $>$ 0 (see \S\S~\ref{adf}).

[{\feii}] lines have been detected in the spectra of some of our sample objects. Unfortunately, these lines 
are severely affected by continuum fluorescence effects \citep[see][]{rodriguez99,verneretal00} and are not suitable for  
abundance determinations. [{\feiii}] lines have been detected in all the objects except at the nucleus of M83. For the calculations of the Fe$^{++}$/H$^{+}$ ratio, we have used a 34-level model atom that uses collision strengths from 
\cite{zhang96} and the transition probabilities of \cite{quinet96} as well as the new transitions found by \cite{johanssonetal00}. The adopted values of the Fe$^{++}$/H$^{+}$ abundance correspond to the mean of the 
abundances obtained for the different [{\feiii}] lines measured for each object. 
Several [{\feiv}] lines have been detected in NGC~2363. The Fe$^{3+}$/H$^{+}$ ratio has been derived using a 33-level model atom where all collision strengths are those calculated by \cite{zhangpradhan97} and the transition probabilities are those recommended by \cite{froesefischeretal08}. 

We have measured several {\hei} emission lines in the spectra of the sample objects. These lines arise mainly from recombination, but they can be affected by collisional excitation and self-absorption effects. We have used the effective recombination coefficients of \citet{storeyhummer95} for {\hi} and those computed by \citet{porteretal05}, with the interpolation formulae provided by \citet{porteretal07} for {\hei}. The collisional contribution was estimated from \citet{saweyberrington93} and \citet{kingdonferland95}, and the optical depth in the triplet lines were derived from the computations by \citet{benjaminetal02}. We have not corrected the {\hei} lines for underlying stellar absorption because: (a) the use of high-spectral resolution data; (b) the lack of appropriate determinations of {\hei} EWs in absorption for OB associations and for all the brightest {\hei} lines along the optical spectrum \citep[see][]{kennicuttetal03}. We have determined the He$^+$/H$^+$ ratio for for {\ts} = 0 and {\ts} $>$ 0 from a maximum likelihood method \citep{peimbertetal00, peimbertetal02}. The {\heii} $\lambda$4686 line has been detected in three objects: NGC~5461, NGC~2363,  
and NGC~4861. We have determined the He$^{++}$/H$^+$ ratio using the line emissivities calculated by \citet{hummerstorey87} and \citet{storeyhummer95}. We have not calculated the He$^{++}$/H$^+$ ratios for {\ts} $>$ 0 considering the small effect of temperature fluctuations on this abundance and the negligible 
contribution of this ion to the total He/H ratio. 

The high signal-to-noise of the spectra has permitted to detect and measure pure RLs of {\oii} and {\cii} in many of the 
sample objects. These lines have almost the same {\elect} and {\elecd} dependence as {\hi} lines. This almost eliminates the bias in the derived 
abundances in the case of line-of-sight temperature variations, with respect to abundances derived from CELs. 

Let $I$($\lambda$) be the intensity of a RL of an element X, $i$ times ionized, at wavelength $\lambda$; then the abundance of the ionization state $i+1$ of element X is given by
 \begin{eqnarray}
  \frac{N({\rm X}^{i+1})}{N({\rm H}^+)} = \frac{\lambda({\rm \AA})}{4861} \frac{\alpha_{eff}({\rm H}\beta)}{\alpha_{eff}(\lambda)} \frac{I(\lambda)}{I({\rm H}\beta)},
 \end{eqnarray}
where $\alpha_{eff}(\lambda)$ and $\alpha_{eff}({\rm H}\beta)$ are the effective recombination coefficients for the line and for H$\beta$, 
respectively. The $\alpha_{eff}({\rm H}\beta)/\alpha_{eff}(\lambda)$ ratio is almost independent of the adopted temperatures and densities.

Following \citet{estebanetal98}, we have determined the O$^{++}$ abundances from the estimated total intensity 
of the RLs of multiplet 1 of {\oii}. This last quantity is obtained by multiplying the sum of the intensities of the individual lines of multiplet 1 of {\oii} observed in a given object by the multiplet correction factor, defined as:   
\begin{eqnarray}
  m_{cf} = \frac{\sum_{all} s_{ij} }{\sum_{obs} s_{ij}},
\end{eqnarray}
where $s_{ij}$ are the theoretical line strengths, which are constructed assuming that they are proportional to the populations of their parent levels, if we 
assume LTE computations predictions. The upper sum runs over $all$ the lines of the multiplet, and the lower sum runs over the $observed$ lines of the multiplet. The lines of multiplet 1 of {\oii} are not in LTE for densities {\elecd} $<$ 10,000 cm$^{\rm -3}$ \citep{ruizetal03,tsamisetal03}. We have used the prescriptions given by \citet{peimbertetal05} to calculate the appropriate 
corrections for the relative strengths between the individual lines of multiplet 1. 
The O$^{++}$ and C$^{++}$ abundances from RLs have been calculated using the representative {\elect}-values of
these ions, {\elect}([{\oiii}]), and the effective recombination coefficients that are available in the literature (\citealt{storey94} for {\oii} assuming LS coupling, and \citealt*{daveyetal00} for {\cii}). In the case of those objects where only upper limits of the {\oii} $\lambda$4549 
intensity have been estimated, the corresponding upper limit of the O$^{++}$ abundance has been derived from the intensity of that line 
and the prescription given by \citet{peimbertetal05} for {\oii} $\lambda$4549.

\subsection{Total Abundances}
\label{totabund} 

To derive the total gas-phase abundances of the different elements present in our 
spectra, we have to correct for the unseen ionization stages by using a set of ionization 
correction factors (ICFs). The total helium abundance has been corrected for the presence of neutral helium using the ionization fractions obtained from the photoionization models of \cite{stasinska90} that reproduce the O abundance, O$^{++}$/O$^+$ ratio, {\elect}([{\oiii}]), and {\elect}([{\nii}]) measured in each object. In 
the case of M83, we have not determined the total He abundance due to the low ionization degree of the nebula,  
that implies a large contribution of neutral helium and a very uncertain ICF. 
In the cases of NGC~5461, NGC~2363, and NGC~4861 --whose spectra show a rather faint  
{\heii} $\lambda$4686 line-- we have assumed that the total He/H ratio is just the sum of 
He$^+$/H$^+$ and a small contribution of He$^{++}$/H$^+$, i.e. the amount of neutral helium is negligible inside 
these nebulae. 

For C we have adopted an ICF-value for the presence of C$^+$ obtained from the photoionization models by  
\cite{garnettetal99}. This correction seems to be fairly appropriate considering the relatively high ionization 
degree of the objects where the {\cii} $\lambda$4267 line has been detected. In order to derive the total abundance of nitrogen we have used the usual ICF based on the similarity between 
the ionization potential of N$^+$ and O$^+$ \citep{peimbertcostero69} in all the cases. Due to the absence of the {\heii} $\lambda$ 4686 line in most of the objects, the total abundance of oxygen has been 
calculated as the sum of O$^+$ and O$^{++}$ abundances. We decided to make the same approximation in the cases of 
NGC~5461, NGC~2363, and NGC~4861 due to the weakness of the {\heii} $\lambda$4686 line detected in these objects, which implies an  
irrelevant contribution of O$^{3+}$. 

The only measurable CELs of Ne in the optical range are those of Ne$^{2+}$ but the fraction of Ne$^{+}$ may 
be important in the nebula. For this element, we have adopted the usual expression proposed by \citet{peimbertcostero69} 
that assumes that the ionization structure of Ne is similar to that of O. We have measured CELs of two ionization stages of S: S$^+$ and S$^{2+}$, and used the ICF proposed by 
\citet{stasinska78} --which is based on photoionization models of \hii\ regions-- to take into account the presence of S$^{3+}$. For argon we have determinations of the Ar$^{2+}$ abundance for all the objects and also of the Ar$^{3+}$ abundance 
for NGC~5461, NGC~5447, NGC~2363, K932, and NGC~4861. However, some contribution of Ar$^{+}$ is expected. We have adopted 
the ICF recommended by \citet{izotovetal94} for this element. 

Finally, for iron, we have determined the Fe$^{++}$ 
abundance for all the objects and also the Fe$^{3+}$ abundance for NGC~2363. Some small contribution of Fe$^{+}$ should be 
present, mainly in the nebulae of lower ionization degree. 
We have used an ICF scheme based on photoionization models of \cite{rodriguezrubin05} to obtain 
the total Fe/H ratio using the Fe$^{++}$ abundances. 

In Tables~\ref{abund1}--\ref{abund3} we show 
the total abundances for {\ts} = 0 and for the {\ts}-value adopted in those objects where this parameter was estimated (see \S\S~\ref{adf}).

\begin{deluxetable}{lcccc} 
\tabletypesize{\scriptsize}
\tablecaption{Ionic and Total Abundances{\TA}
\label{abund3}}
\tablewidth{0pt}
\tablehead{
 & \colhead{NGC~1741} & \colhead{NGC~4395-70} & \colhead{NGC~4861} &  \colhead{M~83}} 
\startdata 
\multicolumn{5}{c}{Ionic Abundances from Collisionally Excited Lines} \\
\hline
N$^+$ & 6.72$\pm$0.09 & 6.35$\pm$0.26 & 5.70$\pm$0.13 & 8.48$\pm$0.09 \\
O$^+$ & 7.77$\pm$0.14 & 7.53$\pm$0.24 & 7.19$\pm$0.24 & 8.94$\pm$0.09 \\
O$^{++}$ & 8.30$\pm$0.07 & 7.95$\pm$0.06 & 7.98$\pm$0.02 & 7.79$\pm$0.12 \\
Ne$^{++}$ & 7.53$\pm$0.10 & 7.24$\pm$0.07 & 7.24$\pm$0.02 & 7.76$\pm$0.24 \\ 
S$^+$ & 5.64$\pm$0.08 & 5.59$\pm$0.22 & 4.89$\pm$0.12 & 6.76$\pm$0.08 \\ 
S$^{++}$ & 6.75$\pm$0.14 & 6.38$\pm$0.11 & 6.23$\pm$0.03 & 6.76$\pm$0.08 \\
Cl$^{++}$ & \nd & 4.32$\pm$0.11 & 4.07$\pm$0.04 & \nd \\ 
Ar$^{++}$ & 5.98$\pm$0.08 & 5.76$\pm$0.06 & 5.60$\pm$0.02 & 6.10$\pm$0.12 \\ 
Ar$^{3+}$ & \nd & \nd & 4.55$\pm$0.02 & \nd \\
Fe$^{++}$ & 5.68$\pm$0.19 & 5.70$\pm$0.30 & 5.18$\pm$0.21 & \nd \\
\hline
\multicolumn{5}{c}{Ionic Abundances from Recombination Lines} \\
\hline
He$^+$ & 10.72$\pm$0.02 & 10.82$\pm$0.02 & 10.90$\pm$0.01 & 10.58$\pm$0.02 \\
He$^{++}$ & \nd & \nd & 8.73$\pm$0.03 & \nd \\
C$^{++}$ & $<$7.97 & $<$7.47 & $<$7.49 & \nd \\ 
O$^{++}$ & $<$8.68 & $<$8.72 &  $<$8.25 & \nd \\
\hline
\multicolumn{5}{c}{Total Abundances} \\
\hline
He & 10.73$\pm$0.02 & 10.83$\pm$0.02 & 10.91$\pm$0.01 & \nd \\
N & 7.36$\pm$0.18 & 6.56$\pm$0.28 & 6.56$\pm$0.28 & 8.51$\pm$0.15 \\
O & 8.41$\pm$0.06 & 8.09$\pm$0.15 & 8.05$\pm$0.04 & 8.97$\pm$0.08 \\
Ne & 7.64$\pm$0.14 & 7.38$\pm$0.17 & 7.30$\pm$0.04 & 8.94$\pm$0.30 \\ 
S & 6.89$\pm$0.13 & 6.51$\pm$0.10 & 6.40$\pm$0.03 & 7.42$\pm$0.23 \\
Ar & 6.23$\pm$0.11 & 5.98$\pm$0.23 & 5.65$\pm$0.02 & \nd \\ 
Fe & 5.95$\pm$0.25 & 5.97$\pm$0.42 & 5.45$\pm$0.33 & \nd \\
\enddata
\tablenotetext{a}{In units of 12+log(X$^{\rm +m}$/H$^+$).}
\end{deluxetable}

\begin{deluxetable}{lccccccccc} 
\tabletypesize{\scriptsize}
\rotate
\tablecaption{Comparison of Ionic Abundances{\TA} and {\ts} Parameter
\label{t2}}
\tablewidth{0pt}
\tablehead{
& \multicolumn{2}{c}{O$^{++}$/H$^+$} & & & \multicolumn{2}{c}{C$^{++}$/H$^+$} & & \\
\colhead{Object} & \colhead{CELs} & \colhead{RLs} & \colhead{ADF(O$^{++}$)} & \colhead{\ts(O$^{++}$)} & \colhead{CELs} & \colhead{RLs} & \colhead{ADF(C$^{++}$)} & \colhead{\ts(C$^{++}$)} & \colhead{\ts(He$^+$)}}
\startdata 
NGC~595 & 7.99$\pm$0.06 & 8.33$\pm$0.07 & 0.34$\pm$0.08 & 0.036$\pm$0.012 & \nd & 8.07$\pm$0.12 & \nd & \nd & 0.033$\pm$0.009 \\
NGC~604 & 8.19$\pm$0.02 & 8.45$\pm$0.04 & 0.26$\pm$0.04 & 0.034$\pm$0.008 & \nd & 8.19$\pm$0.11 & \nd & \nd & 0.022$\pm$0.010 \\
H1013 & 8.06$\pm$0.12 & 8.42$\pm$0.11 & 0.36$\pm$0.13 & 0.037$\pm$0.020 & \nd & 8.40$\pm$0.12 & \nd & \nd & 0.031$\pm$0.017 \\
NGC~5461 & 8.29$\pm$0.03 & 8.38$\pm$0.07 & 0.09$\pm$0.08 & 0.014$\pm$0.014 & 8.27{\TB}/7.90{\TC} & 8.13$\pm$0.20 & $-$0.14/0.23 & $<$0/0.028 & 0.057$\pm$0.018 \\
NGC~5447 & 8.27$\pm$0.02 & 8.46$\pm$0.07 & 0.19$\pm$0.07 & 0.032$\pm$0.007 & \nd & 8.06$\pm$0.11 & \nd & \nd & 0.094$\pm$0.034 \\
VS~24 & 8.05$\pm$0.06 & $<$8.50 & $<$0.45 & \nd & \nd & 8.15$\pm$0.24 & \nd & \nd & 0.075$\pm$0.012 \\
VS~38 & 7.93$\pm$0.04 & $<$8.54 & $<$0.61 & \nd & 8.00{\TB}/7.84{\TC} & 8.27$\pm$0.15 & 0.27/0.43 & 0.026/0.035 & 0.057$\pm$0.015 \\
VS~44 & 8.13$\pm$0.03 & 8.43$\pm$0.06 & 0.30$\pm$0.06 & 0.039$\pm$0.010 & 8.08{\TB}/7.85{\TC} & 7.98$\pm$0.18 & $-$0.10/0.13 & $<$0/0.018 & 0.066$\pm$0.010 \\
NGC~2363 & 7.73$\pm$0.02 & 8.02$\pm$0.05 & 0.29$\pm$0.05 & 0.120$\pm$0.022 & 7.15{\TD} & 7.55$\pm$0.09 & 0.40 & 0.153 & 0.101$\pm$0.019 \\
K932 & 8.28$\pm$0.02 & 8.52$\pm$0.04 & 0.24$\pm$0.04 & 0.033$\pm$0.007 & \nd & 8.31$\pm$0.13 & \nd & \nd & 0.030$\pm$0.009 \\
NGC~1741 & 8.30$\pm$0.07 & $<$8.68 & $<$0.38 & \nd & \nd & $<$7.97 & \nd & \nd & 0.035$\pm$0.076 \\
NGC~4395-70 & 7.95$\pm$0.06 & $<$8.72 & $<$0.77 & \nd & \nd & $<$7.47 & \nd & \nd & 0.125$\pm$0.064 \\
NGC~4861 & 7.98$\pm$0.02 & $<$8.25 & $<$0.27 & \nd & 7.44$\pm$0.11{\TE} & $<$7.49 & $<$0.05 & \nd & 0.103$\pm$0.020 \\
M~83 & 7.79$\pm$0.12 & \nd & \nd & \nd & \nd & \nd & \nd & \nd & 0.036$\pm$0.010 \\
\enddata
\tablenotetext{a}{In units of 12+log(X$^{\rm +m}$/H$^+$).}
\tablenotetext{b}{Garnett et al. (1999), assuming $R_{\rm V}$ = 3.1.}
\tablenotetext{c}{Garnett et al. (1999), assuming $R_{\rm V}$ = 5.}
\tablenotetext{d}{Garnett et al. (1995).}
\tablenotetext{e}{Kobulnicky \& Skillman (1998).}
\end{deluxetable}

\section{Discussion}
\label{discussion} 

\subsection{The Abundance Discrepancy Factor in the Sample Objects}
\label{adf} 

\begin{deluxetable}{lcccc} 
\tabletypesize{\scriptsize}
\tablecaption{Abundance Gradients in M33, M101, NGC~2403, and the Milky Way
\label{gradients}}
\tablewidth{0pt}
\tablehead{
Galaxy & O/H (CELs) & O/H (RLs) & C/H & C/O (RLs)} 
\startdata 
\multicolumn{5}{c}{With respect to R$_{\rm G}${\TA}} \\
\hline
M~33 & $-$0.056 $\pm$ 0.029 & $-$0.072 $\pm$ 0.047 & $-$0.105 $\pm$ 0.120 &  -0.033 $\pm$ 0.130 \\
M~101 & $-$0.022 $\pm$ 0.004 & $-$0.017 $\pm$ 0.009 & $-$0.046 $\pm$ 0.012 &  -0.029 $\pm$ 0.020 \\
NGC~2403 & $+$0.012 $\pm$ 0.021 & \nd & $-$0.097 $\pm$ 0.127 & \nd \\
Milky Way{\TB} & $-$0.040 $\pm$ 0.006 & $-$0.044 $\pm$ 0.010 & $-$0.103 $\pm$ 0.018 &  -0.059 $\pm$ 0.020 \\
\hline
\multicolumn{5}{c}{With respect to R$_{\rm 0}${\TC}} \\
\hline
M~33 & $-$0.39 $\pm$ 0.20 & $-$0.50 $\pm$ 0.32 & $-$0.72 $\pm$ 0.83 &  -0.22 $\pm$ 0.98 \\
M~101 & $-$0.62 $\pm$ 0.10 & $-$0.50 $\pm$ 0.25 & $-$1.32 $\pm$ 0.33 &  -0.65 $\pm$ 0.49 \\
NGC~2403 & $+$0.09 $\pm$ 0.17 & \nd & $-$0.77 $\pm$ 1.00 & \nd \\
\enddata
\tablenotetext{a}{In units of dex kpc$^{-1}$.}
\tablenotetext{b}{Esteban et al (2005).}
\tablenotetext{c}{In units of dex R$_{\rm 0}^{-1}$.}
\end{deluxetable}

In Table~\ref{t2}, we include the O$^{++}$ and C$^{++}$ abundances determined from RLs and CELs for the sample 
objects as well as the ADF for each ion. The C$^{++}$ abundance determinations based on CELs included in the table have been  calculated from the intensities of 
[\ciii] $\lambda$1907 + \ciii] $\lambda$1909 lines, which are in the UV and come from the literature 
\citep{garnettetal95,garnettetal99,kobulnickyskillman98}. Except in the cases of NGC~2363 and NGC~4861, the rest of the C$^{++}$/H$^+$ ratios based 
on CELs have two possible values, which correspond to the higher and lower limits adopted by \citet{garnettetal99} because 
of the uncertainty in the choice of the UV reddening function from their data. \citet{garnettetal95} determined the 
C$^{++}$ abundance of NGC~2363 from UV CELs, and because of its low reddening, the uncertainty in the 
 UV reddening function is not a concern for this object. In the case of NGC~4861, the choice of the UV reddening law can also affect 
 but not very importantly because its relatively low extinction \citep[see][]{kobulnickyskillman98}.

As it can be seen in Table~\ref{t2}, the values of the ADF(O$^{++}$) are remarkably similar in all the objects, with a mean 
value of 0.26 $\pm$ 0.09 dex, which is very similar to the mean ADF(O$^{++}$) obtained 
for a sample of Galactic \hii\ regions \citep[see][and references therein]{garcia-rojasesteban07} and 
for other EHRs \citep[0.30 $\pm$ 0.09 dex, see][and references therein]{lopez-sanchezetal07}. The upper limits of the 
ADF(O$^{++}$) included in Table~\ref{t2} are always of the order or larger than the mean ADF(O$^{++}$). In the case of the ADF(C$^{++}$) 
the results 
are not so clear because of the limited number of determinations 
available in the literature, aperture effects, and the uncertainty in the UV reddening function assumed to derive the abundance from UV CELs, 
however, our values of the ADF(C$^{++}$) for NGC~2363 and those corresponding 
to $R_V$ = 5.0 for NGC~5461, VS~38, and VS~44 are fairly similar to the ADF(C$^{++}$)-values compiled by 
\citet{lopez-sanchezetal07} for other Galactic and extragalactic \hii\ regions. In the case of NGC~4861 we find a puzzling result because its 
upper limit of the ADF(C$^{++}$) is about zero.  

As a conclusion of all the results about the ADF-values gathered in this and previous papers by our group, the ADF  
seems to be a remarkably constant quantity in \hii\ regions. It seems clear that whatever phenomenon produces this discrepancy, it does not depend on the metallicity of the \hii\ region, nor on its galactocentric distance --when the region belongs to a spiral galaxy-- nor on the morphological type (or the absolute magnitude) of the host galaxy. This is a fairly different behavior from the one shown by the ADF in PNe, where this parameter  
changes almost by one order of magnitude from one object to the other. The results of this paper reinforce the suggestion of \citet{garcia-rojasesteban07} that the physical mechanism that produces the abundance discrepancy --or the bulk of it-- 
in the PNe with large ADF-values should be different to that acting in \hii\ regions. 

Assuming the validity of the temperature fluctuations paradigm and that this phenomenon produces the abundance discrepancy, we 
can estimate the values of the \ts\ parameter for each object. In Table~\ref{t2}, we include the 
\ts\--values that produce the agreement between the abundance determinations obtained from CELs and RLs of O$^{++}$ and C$^{++}$.  These calculations have been made following the formalism outlined by \cite{peimbertcostero69}. As we have indicated along the paper, the ionic abundances, ICFs, and total abundances have been calculated considering both possibilities: \ts\ = 0 and the \ts-values 
obtained from the 
ADF(O$^{++}$). We have assumed that the same \ts\ is valid for determining the abundances of the different ionic species. 
As we expect from the similarity of the ADF(O$^{++}$) found for the sample objects, the corresponding values of the \ts\ parameter obtained are also fairly similar, and also similar to those obtained for Galactic \hii\ regions and other EHRs 
\citep[see][]{garcia-rojasesteban07,lopez-sanchezetal07}. It is also interesting to note that the \ts\--value determined from the ADF(C$^{++}$) 
for NGC~2363 is consistent with that determined from the ADF(O$^{++}$). For the rest of the sample objects where the ADF(C$^{++}$) has been 
estimated, a firm conclusion cannot be drawn due to the strong dependence of the associated \ts\--value on the assumed UV reddening law. In Table~\ref{t2}, we also include the \ts\--values obtained from the application of a maximum likelihood method \citep{peimbertetal00, peimbertetal02} to 
search for the physical conditions, He$^+$/H$^+$ ratios, and optical depths that would be a simultaneous 
fit to the measured lines of {\hei} (see \S\S~\ref{ionabund}). As we can see in Table~\ref{t2}, the values of  {\ts}(He$^+$) and {\ts}(O$^{++}$) are consistent within the errors in the cases of NGC~595, H1013, NGC~2363, and  
K932 (50\% of the objects where both quantities were determined), marginally consistent in the case of NGC~604, and in disagreement for NGC~5461, NGC~5447, and VS~44. 

The results included in Table~\ref{t2} are fairly consistent with previous determinations of the \ts\ parameter for some of the 
sample objects. \citet{estebanetal02} obtained \ts-values of 0.027 $\pm$ 0.018, 0.041 $\pm$ 0.021, and 0.128 $\pm$ 0.045 for 
NGC~604, NGC~5461, and NGC~2363, respectively, also from the ADF(O$^{++}$). In the case of NGC~2363, \citet{estebanetal02} obtain 
\ts\ $\sim$ 0.102 from the ADF(C$^{++}$). Interestingly, \citet{gonzalez-delgadoetal94} estimate a \ts\ = 0.064 $-$ 0.098 
from the independent method based on the comparison of {\elect}([{\oiii}]) and $T$(Pac) (Paschen continuum) for NGC~2363. 
In the case of NGC~2363, it is remarkable that four independent methods for estimating \ts\ give entirely consistent results. 
Finally, \citet{bresolin07} obtains a 
\ts\ = 0.06 $\pm$ 0.02 for H~1013 from the comparison of {\elect}([{\oiii}]) and $T$(Bac) (Balmer continuum). 

Clearly, the fact that RLs and CELs provide abundances that differ by about 0.3 dex is an important concern in nebular astrophysics. An additional problem is that we do not know which kind of lines gives the correct abundances. This situation means that the 
chemical composition of cosmic objects derived from \hii\ regions are in a sort of quarantine and perhaps 
should be revised. This puzzling situation has important 
implications for different relevant astrophysical issues such as: (a) the ingredients of chemical evolution models and predicted stellar yields \citep[see][]{carigietal05}; (b) the luminosity- and mass-metallicity relations for local and high-redshift star-forming galaxies \citep[see][]{salzeretal05}; (c) the calibration of strong-line methods for deriving \hii\ region abundances \citep{peimbertetal07}; 
(d) the possible metallicity dependence of the Cepheid period-luminosity relation \citep{sakaietal04}; (e) the metallicity dependence of the number ratios of the different types of Wolf-Rayet stars \citep{meynetmaeder05}; (f) the comparison of the Solar abundances with the Orion Nebula abundances \citep{carigipeimbert08}.

\subsection{The C/H and C/O Radial Gradients in M~33, M~101, and NGC~2403}
\label{gradients} 
\begin{figure*}
 \includegraphics[angle=0,scale=1]{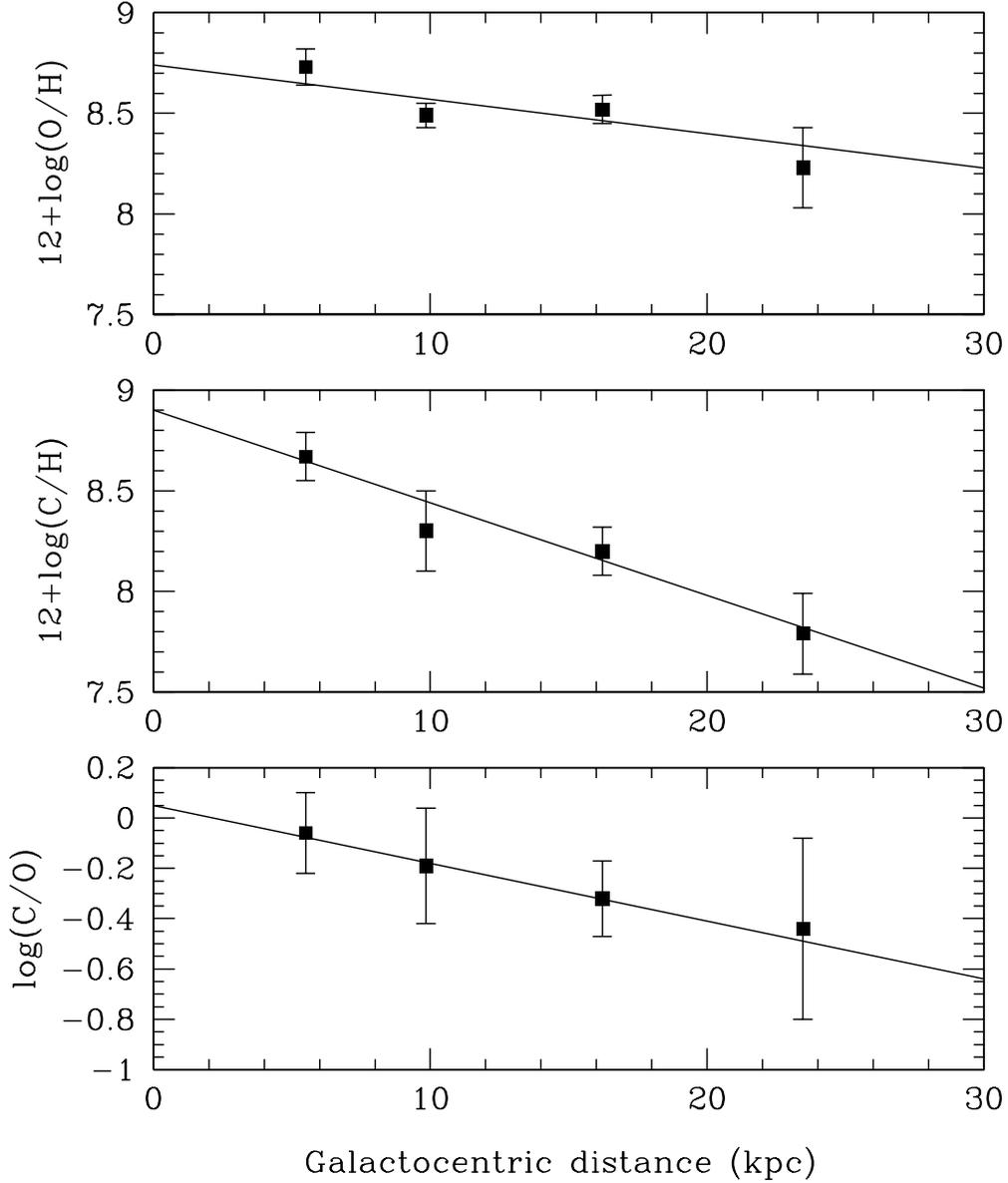} 
  \caption{O, C, and C/O radial abundance gradients of the disk of the spiral galaxy M~101 from \hii\ region abundances determined from recombination lines. The lines indicate the least-squares linear fits to the data. The point at $R_G$ = 23.45 kpc corresponds to NGC~5471 and has been taken from \cite{estebanetal02}. }
  \label{grad_M101}
\end{figure*}

The determination of the O/H and/or C/H ratios from RLs in two or more \hii\ regions in the spiral galaxies M~33, M~101, and NGC~2403 allows us to estimate the 
radial abundance gradients of those two elements in these galaxies. Considering the O and C abundances determined in this paper and the galactocentric 
distances and $R$/$R_{0}$ ratios compiled in Table~\ref{observations}, we have obtained the O/H (CELs), O/H (RLs), C/H (RLs) and C/O (RLs) gradients shown in Table~\ref{gradients}. In the case of M~101, we have also included the data for NGC~5471 obtained by \citet{estebanetal02} for the gradient determinations. In Figure~\ref{grad_M101}, we show the gradients found for M~101. It is remarkable the clear trend of the gradients found for this 
galaxy. 
For comparison, in Table~\ref{gradients} we also include the C and O gradients determined by \citet{estebanetal05} for the Milky Way. An important  
result in all the cases is that the slope of the O abundance gradients is very similar independently of the kind of line --RLs or CELs-- used for the 
abundance calculation. This is a likely consequence of the lack of correlation between the ADF(O$^{++}$) and metallicity.  

The O abundance gradients determined from CELs of these galaxies have been studied in detail by many authors, especially for M~33 and M~101. The values of the O/H ratio that we obtain from CELs for each individual \hii\ region of these galaxies do not 
differ typically more than 0.1 or 0.2 dex from the values found in the literature. The largest differences are for 
VS~24 and VS~38 of NGC~2403. The O abundances found by \cite{garnettetal97} are higher than ours by a factor of 0.39 dex and 
0.24 dex for VS~24 and VS~38, respectively, this can be due to the more uncertain data of those authors. In fact, \cite{garnettetal97} find a 
negative reddening coefficient for VS~24 --we obtain $c$(H$\beta$) = 0.34, a value more similar to those found for the 
rest of the \hii\ regions observed in NGC~2403-- and they do not obtain a direct determination of {\elect}([{\oiii}]) 
for VS~38.

In the case of M~33, we obtain a slope of the O/H gradient that is flatter than that of \citet{vilchezetal88} ($-$0.12 $\pm$ 0.01 dex kpc$^{-1}$), steeper than those determined by \citet{crockettetal06} and \citet{rosolowskysimon08} (values of $-$0.01 $\pm$ 0.01 and  $-$0.03 $\pm$ 0.01 dex kpc$^{-1}$, respectively), but similar to those obtained by \citet{urbanejaetal05} --from the
spectral analysis of B-type supergiant stars-- and \citet{magrinietal07}. For M~101, our O/H gradient 
is consistent with that of $-$0.029 $\pm$ 0.003 dex kpc$^{-1}$ obtained by \citet{kennicuttetal03} and slightly flatter than 
that obtained by \citet{estebanetal02} from RLs of only two \hii\ regions of the galaxy. However, our C/O gradient is 
somewhat steeper than the estimate of \citet{estebanetal02} but consistent with the value determined by \citet{garnettetal99} from CELs, which  
is between $-$0.025 and $-$0.040, depending on the adopted UV reddening function.
In the case of NGC~2403, our O/H (CELs) gradient is slightly positive but with a large uncertainty, in contrast to the negative gradients obtained by \citet{fierroetal86}, \citet{garnettetal97}, and \citet{vanzeeetal98}, who find values between $-$0.07 and $-$0.10. This result could be spurious and 
due to the narrow baseline of galactocentric distances covered with the three \hii\ 
regions that we have observed in this galaxy. 

As \citet{estebanetal05} found for the Milky Way, the C/H gradients are steeper than the ones obtained for the O/H ratio in all the galaxies, producing the negative slopes of the C/O gradients. Based on chemical evolution models of the Milky Way, \citet{carigietal05} have found that such values of the slope of the C/O gradient can only be reproduced when 
carbon yields that increase with metallicity owing to stellar winds in massive stars are considered \citep[as the yields obtained 
by][]{maeder92,meynetmaeder02,hirschietal05}. However, it is clear that taylored chemical evolution models for each individual spiral galaxy would be necessary in order to reach more precise conclusions on this issue.   

\begin{figure*}
  \includegraphics[angle=0,scale=1]{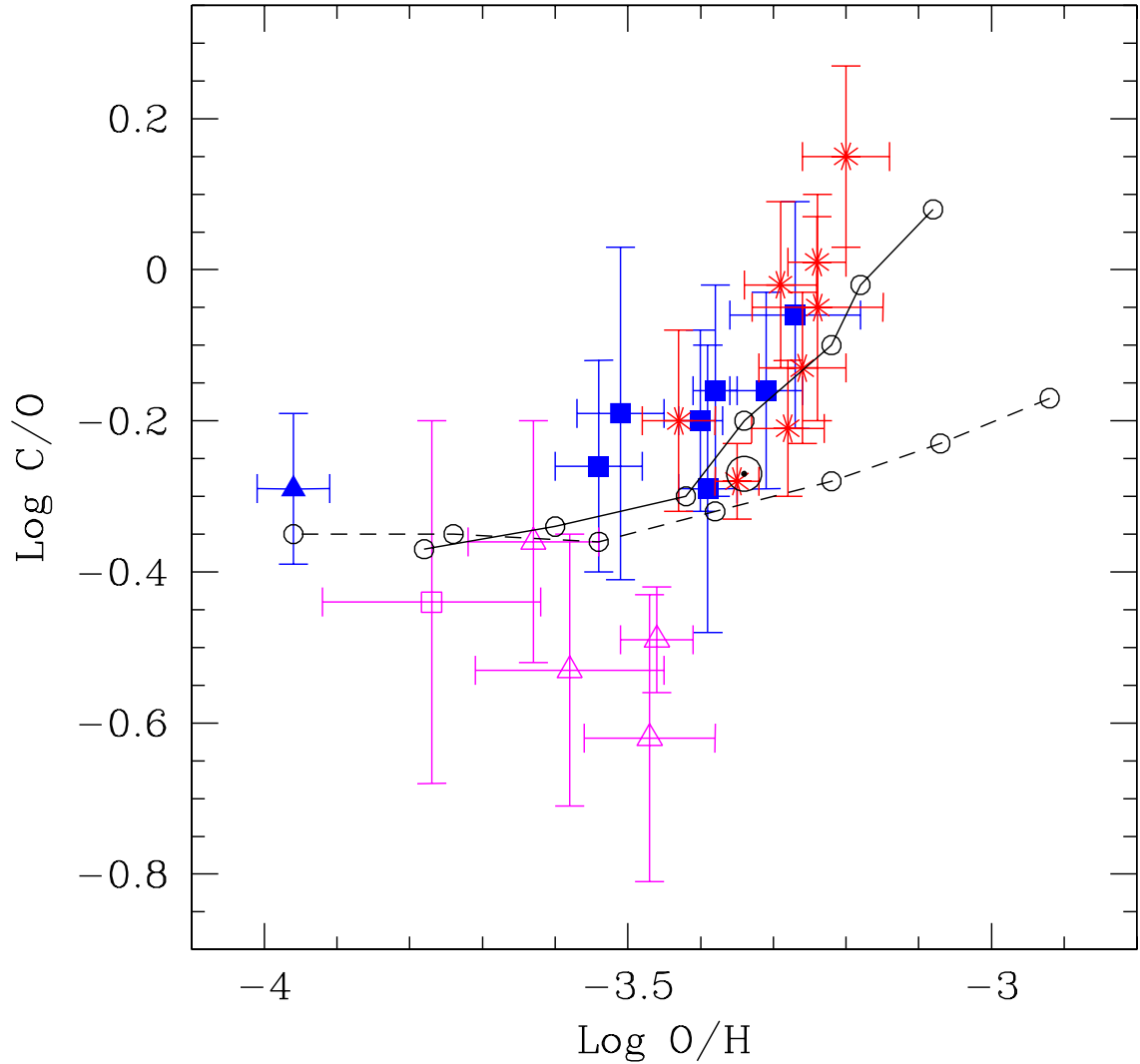} 
  \caption{C/O vs. O/H ratios of Galactic and extragalactic \hii\ regions (EHRs) determined from recombination lines. Filled (blue) symbols represent 
  the data analysed in this paper; the filled squares correspond to abundance ratios of EHRs in spiral galaxies and the filled triangle 
  to NGC~2363, an EHR in a dwarf galaxy. The empty (magenta) square corresponds to NGC~5471, an EHR 
at the outskirts of M~101 \citep[data from][]{estebanetal02}. Empty (magenta) triangles correspond to data for EHRs in dwarf galaxies taken from the literature: 30 Dor \citep{peimbert03}, region V of NGC~6822 \citep{peimbertetal05}, and regions A and C of NGC~5253 \citep{lopez-sanchezetal07}. The (red) asterisks 
  correspond to data for Galactic \hii\ regions \citep{garcia-rojas06,garcia-rojasesteban07}. The solar symbol (large circle)  
  represents the abundances of the Sun \citep{grevesseetal07}. The lines show the predictions of chemical evolution models by \cite{carigietal05} and \cite{carigipeimbert08} for the present-day Galactic disk at different Galactocentric distances for high-wind yields and M$_{up}$ = 80 M$_{\odot}$ (solid line) and for low-wind yields and M$_{up}$ = 80 M$_{\odot}$ (dashed line). In both lines, the small open circles represent the predictions of the models at different Galactocentric distances, from right to left: 4, 6, 8, 10, 12, 14, and 16 kpc.}
  \label{co_vs_o_RG}
\end{figure*}

A general correlation between C/O and O/H has been obtained by \citet{garnettetal95} and \citet{garnettetal99} from observations of CELs in 
dwarf irregular and spiral galaxies and by \citet{estebanetal02} from observations of RLs in a small sample of Galactic and extragalactic 
\hii\ regions. In Figure~\ref{co_vs_o_RG} we have included the results for the EHRs observed in this paper and others from the literature 
\citep{estebanetal02,peimbert03,peimbertetal05,lopez-sanchezetal07} as well as the points corresponding to the sample of Galactic \hii\ regions of \citet{garcia-rojas06} and \citet{garcia-rojasesteban07}. 
As we can see in the figure, the Galactic objects and the \hii\ regions belonging to spiral galaxies follow a rather clear trend, where the C/O ratio experiences a strong increase for O abundances between 8.5 and 8.8. This behaviour has also been observed in 
samples of nearby F and G dwarf stars of the Galactic disk \citep{bensbyfeltzing06} and dwarf and subgiant stars of the same spectral types 
belonging to the Galactic halo population \citep{akermanetal04}. \citet{bensbyfeltzing06} remark that the C/O vs. O/H trend is fairly similar to the well-known Fe/O vs. O/H one, suggesting that the low- and intermediate-mass stars should be important contributors 
to the C enrichment at higher metallicities. This was also one of the main results of the chemical evolution models 
by \citet{carigietal05}. 

In Figure~\ref{co_vs_o_RG}, we include the predictions of chemical evolution models for the present-day Galactic disk at different Galactocentric distances \citep{carigipeimbert08} that reproduce the observed C, O, and C/O gradients of the Galactic disk determined by \citet{estebanetal05}. The solid line represents the results of a model that assumes the high-wind yields by \citet{maeder92} and M$_{up}$ = 80 M$_{\odot}$ \citep{carigietal05}. The dashed line represents the results of a model that considers the low-wind yields by \citet{hirschietal05} and M$_{up}$ = 80 M$_{\odot}$ \citep{carigipeimbert08}. In Figure~\ref{co_vs_o_RG}, we can see 
that the predictions for high-wind yields match quite well the distribution of the points corresponding to the \hii\ regions in the Milky Way and other spiral galaxies. This is an important result considering that the objects 
belong to spiral galaxies of different morphological type than the Milky Way, different masses, and different absolute magnitudes. 
This would imply that the assumptions used to generate the chemical evolution model of our Galaxy are reasonable 
for other spirals as well.  
In general, the O/H and C/O ratios of \hii\ regions in dwarf galaxies are lower than those of the objects in spiral galaxies, and their C/O 
ratios also show a larger scatter. \citet{garnettetal99} indicated that changes in the assumed star formation law through its dependence on the power of 
the gas surface density can change the curves of the C/O vs. O/H relation in the C/O axis. This can be the case of dwarf galaxies with respect 
to the more massive spiral galaxies, since their gas consumption timescales must be very different and models for the Milky Way should not 
necessarily reproduce their chemical evolution.

The O and C abundances of the \hii\ regions presented in Figure~\ref{co_vs_o_RG} have not been corrected for the fraction of atoms embedded
in dust, the Fe/O ratio in the observed \hii\ regions is at least one order of magnitude smaller than in the Sun implying that most
of the Fe is trapped in dust grains. According to \cite{estebanetal98} and to \cite{mesadelgadoetal09} the fraction of O and C
embedded in dust grains in the Orion Nebula is about 0.1 dex. By assuming that some fraction of C and O is embedded in dust grains, 
the gas plus dust C/O values of the \hii\ regions in Figure~\ref{co_vs_o_RG} would not be affected while the gas
plus dust O/H values should be increased. An increase of the O/H ratios slightly lower than that estimated for the Orion Nebula 
(i.e. 0.05 dex) would produce a better agreement between the observations and the chemical evolution models with high-wind yields 
presented in this figure. At any rate, strong conclusions cannot be drawn from this comparison considering the large uncertainties in both 
the chemical evolution models and observations.  

\section{Conclusions}
\label{conclu}

Echelle spectrophotometry from 3550 to 7440 \AA\ has been obtained with the HIRES spectrograph at the Keck I 
telescope for 13 extragalactic \hii\ regions (EHRs). One additional object was observed with the WHT telescope at lower spectral 
resolution. The EHRs observed include objects in the spiral galaxies M~31, M~33, M~83, M~101, NGC~2403, and NGC~4395 as well as in 
irregular and blue compact dwarf galaxies: NGC~1741, NGC~2366, and NGC~4861. 

The spectra are very deep and a number of lines between 38 and 135 have been measured in each EHR. We detected \cii\ lines 
in 10 of the objects and \oii\ lines in 8 of them. This allows us to derive the C$^{++}$ and O$^{++}$ abundances 
from recombination lines, which are independent of possible variations of the spatial distribution of the electron temperature 
in the nebulae. 

We have obtained a large number of different {\elect} determinations, in particular those associated with low ionization 
potential ions: [{\nii}], [{\oii}], and [{\sii}]. We have obtained a rather tight empirical linear relation between {\elect}([{\nii}]) and {\elect}([{\oiii}]), which is almost identical to the relationship found by \citet{garnett92} from photoionization models. The use of 
this relation is of interest for abundance determinations of  
\hii\ regions where {\elect}([{\oiii}]) can not be determined. A linear relationship is also found between {\elect}([{\oii}]) and {\elect}([{\oiii}]) and between {\elect}([{\sii}]) and {\elect}([{\oiii}]), but with a larger scatter in this last case. 

The O$^{++}$ abundance determined from \oii\ recombination lines is always larger than that determined from collisionally 
excited lines of [{\oiii}] in all the objects where both quantities have been calculated. The mean difference between both 
abundances --the so-called abundance discrepancy factor (ADF)-- is 0.26 $\pm$ 0.09 dex --about a factor of 2, which  
is also similar to the typical values of this parameter in Galactic \hii\ regions and EHRs from the literature. This result 
indicates that whatever mechanism is producing the abundance discrepancy in \hii\ regions, it seems independent 
of the properties of the \hii\ regions and of the parent galaxy. 

We have determined the O/H, C/H, and C/O radial gradients in the spiral galaxies: M~33, M~101, and NGC~2403. We find 
that the C/H gradients are always steeper than those of O/H, producing negative slopes of the C/O gradient. The behavior 
of these gradients is similar to that of the Milky Way determined by \citet{estebanetal05}, and can be reproduced 
when carbon yields that increase with metallicity owing to stellar winds in massive stars are considered \citep{carigietal05,carigipeimbert08}. 
The relation between C/O and O/H in extragalactic and Galactic \hii\ regions shows a rather clear trend, where the C/O 
ratio shows a strong increase for O abundances between 8.5 and 8.8, indicating that the low- and intermediate-mass stars 
should be important contributors to the C enrichment at higher metallicities.                                                           

We are grateful to the referee of this paper for his/her useful comments. 
This work has been funded by the Spanish Ministerio de Ciencia y Tecnología (MCyT) under project AYA2004-07466,  Ministerio de Educaci\'on y Ciencia (MEC) under project AYA2007-63030, and the Mexican CONACyT grant 46904. 
F.B gratefully acknowledges the support from the NSF grant AST-0707911.

\end{document}